\documentclass[aps,prb,twocolumn,superscriptaddress,floatfix,showpacs,amsmath,amssymb,nofootinbib,longbibliography,nobibnotes,noeprint]{revtex4-2}

\usepackage{graphicx}% Include figure files
\usepackage{dcolumn}% Align table columns on decimal point
\usepackage{bm}% bold math
\usepackage{color}
\usepackage{tabularx}
\usepackage{csquotes}
\usepackage[normalem]{ulem}
\usepackage{xcolor}
\usepackage{makecell}
\usepackage{nicefrac}
\usepackage{soul}
\usepackage{array}
\usepackage{multirow}

\newcommand{\abx}{X$_3$BA}
\newcommand{\LiMChO}{(Li$_2M$)$Ch$O}
\newcommand{\LiMnChO}{(Li$_2$Mn)$Ch$O}
\newcommand{\LiFeChO}{(Li$_2$Fe)$Ch$O}
\newcommand{\LiCoChO}{(Li$_2$Co)$Ch$O}
\newcommand{\LiMSeO}{(Li$_2M$)SeO}
\newcommand{\LiMSO}{(Li$_2M$)SO}
\newcommand{\LiFeSeO}{(Li$_2$Fe)SeO}

\newcommand{\LiFeSO}{(Li$_2$Fe)SO}

\newcommand{\LiMnSeO}{(Li$_2$Mn)SeO}
\newcommand{\LiMnSO}{(Li$_2$Mn)SO}
\newcommand{\LiCoSeO}{(Li$_2$Co)SeO}
\newcommand{\LiCoSO}{(Li$_2$Co)SO}

\newcommand{\chifc}{$\chi_{\rm FC}$}
\newcommand{\chizfc}{$\chi_{\rm ZFC}$}
\newcommand{\ergGmol}{erg/(G$^2$mol)}

\newcommand{\fezn}{Fe$_{1-x}$Zn$_x$F$_2$}
\newcommand{\mnf}{MnF$_2$}

\newcommand{\comg}{Co$_{1-x}$Mg$_x$O}

\newcommand{\rev}{\color{black}}

\newcommand{\rk}{\color{black}}

\newcommand{\lb}{\color{purple}}

\newcommand{\etal}       {{\it et~al}.}
\newcommand{\lifeso}    {$({\mathrm{Li}}_{2} {\mathrm{Fe}}) {\mathrm{S}} {\mathrm{O}}$}
\newcommand{\lifeseo}   {$({\mathrm{Li}}_{2} {\mathrm{Fe}}) {\mathrm{Se}} {\mathrm{O}}$}
\newcommand{\tn}      {$T_{\rm N}$}

\newcommand{\tc}      {$T_{\rm C}$}
\newcommand{\tirr}      {$T_{\rm irr}$}

\newcommand{\mb}    {$\mu_{\mathrm{B}}$}
\newcommand{\tleft}{$T_{\mathrm{l}}$}
\newcommand{\tright}{$T_{\mathrm{h}}$}

\newcommand{\degree}{$\,^{\circ}$}

\begin{document}

%\preprint{APS/123-QED}

\title{{\rk Magnetism in antiperovskite (Li$_2$\textit{M})\textit{Ch}O (\textit{M} = Fe, Mn, Co; \textit{Ch} = S, Se) diluted magnets with fixed 1/3 filling: the key role of magnetic anisotropy}}

%%%%  AUTHORS %%%%%%%%%%%%%%%
\author{J.~Zheng}
\affiliation{Kirchhoff Institute for Physics, Heidelberg University, INF 227, D-69120 Heidelberg, Germany}

\author{F.~L. Carstens}
\affiliation{Kirchhoff Institute for Physics, Heidelberg University, INF 227, D-69120 Heidelberg, Germany}

\author{L.~Singer}
\affiliation{Kirchhoff Institute for Physics, Heidelberg University, INF 227, D-69120 Heidelberg, Germany}

\author{M.A.A. Mohamed} \affiliation{Leibniz Institute for Solid State and Materials Research Dresden, 01069 Dresden, Germany}
\affiliation{Department of Physics, Faculty of Science, Sohag University, 82524 Sohag, Egypt}

\author{{\rk L.~Bischof}}
\affiliation{Kirchhoff Institute for Physics, Heidelberg University, INF 227, D-69120 Heidelberg, Germany}

\author{{\rk A.~Alfonsov}} \affiliation{Leibniz Institute for Solid State and Materials Research Dresden, 01069 Dresden, Germany}

\author{J.~Arneth}
\affiliation{Kirchhoff Institute for Physics, Heidelberg University, INF 227, D-69120 Heidelberg, Germany}

\author{S. Hampel} \affiliation{Leibniz Institute for Solid State and Materials Research Dresden, 01069 Dresden, Germany}

\author{N.~Gräßler} \affiliation{Leibniz Institute for Solid State and Materials Research Dresden, 01069 Dresden, Germany}

\author{R.~Klingeler}
\email{klingeler@kip.uni-heidelberg.de}
\affiliation{Kirchhoff Institute for Physics, Heidelberg University, INF 227, D-69120 Heidelberg, Germany}

\date{\today}

\begin{abstract}
We report the magnetic properties of a series of lithium-rich antiperovskites \LiMChO\ ($M$ = Fe, Co, Mn and $Ch$ = Se, S) where transition metal and lithium ions are randomly distributed on the X-sites of the \abx\ structure, thereby forming a strongly diluted magnetic sublattice. Our study hence enables us to investigate the evolution of magnetic order at fixed 1/3-filling -- which is in the vicinity but slightly above the percolation threshold -- upon variation of the spin size, the magnetic anisotropy, and the orbital configuration. The data imply the absence of a distinct Curie-Weiss behavior up to 350~K but show rather large and weakly temperature-dependent magnetic susceptibility. We observe clear signatures of long-range antiferromagnetic order evolving in the 1/3-filled and strongly diluted magnetic X-site lattice with increasing N\'eel temperatures from $T_{\rm{N}}\simeq 30$~K in \LiMnChO\ to $\simeq 50$~K in \LiFeChO\ and $70-90$~K in \LiCoChO . Except for $M$ = Co, the chalcogenide has no sizable effect on $T_{\rm N}$. {\rk We conclude significant magnetic coupling and short-range magnetic correlations at well above \tn\ which is in line with the observation of a broad electron spin resonance signal at room temperature}. The actual ordering temperatures are strongly diminished by magnetic dilution. While structural parameters such as the tolerance factor and bonding angles do not strongly affect $T_{\rm N}$, a key parameter is the magnetic anisotropy of the transition metals.
\end{abstract}
\maketitle

%%%%%%%%%%%%%%%%%%%%%%%%%%
\section{Introduction}
%%%%%%%%%%%%%%%%%%%%%%%%%%

In diluted magnetic systems, a fraction of the magnetic ions occupying the crystallographic lattice sites is substituted by non-magnetic species. The random distribution of magnetic and non-magnetic ions reduces the number of exchange interaction pathways and introduces significant disorder. This leads to the suppression of long-range magnetic order and the emergence of novel magnetic phases and  phenomena.~\cite{khairnar2021,Birgeneau1984,Sousa2010,Kannan1986} Examples include glass-like magnetism and Griffiths phases, which originate from inhomogeneous magnetic exchange interactions in rare regions and give rise to power-law critical behavior.~\cite{majee2019,gupta2017,binek1995,lu2024,shand1998,Montenegro1988} This is particularly evident at dilution levels near the percolation threshold of the magnetic sublattice or in the presence of local geometric frustration.~\cite{Birgeneau1984,martinho2001} A variety of unconventional magnetic properties may be associated with the (partly) fragmentation of the percolating magnetic network and the formation of finite-sized magnetic clusters. When the  concentration of magnetic ions is sufficiently low and falls below the percolation limit, the systems do not exhibit long-range magnetic order. Above the threshold, the magnetic ordering temperature can be strongly suppressed compared to non-diluted systems and the associated anomalies in the thermodynamic response functions are often weak and broad.~\cite{Sousa2010,Kannan1986,kemei2014}

The recently discovered family material of Li-rich antiperovskites \LiMChO\ ($M$ = Fe, Co, Mn and $Ch$ = Se, S) adds a new branch to this field~\cite{lai2017,lu2018}. The material resembles a cubic structure with space group of $Pm\overline{3}m$ where diamagnetic Li$^+$ ions and paramagnetic $M^{2+}$ ions are statistically distributed at the same crystallographic site (3$c$), i.e., X in the antiperovskite \abx\ structure. The arrangement of metal ions may be considered as an incomplete fcc lattice formed by the face-centered sites of the fcc structure while omitting the corner ones. The magnetic 3$c$ site in the antiperovskite structure is filled with 1/3 magnetic ions % which is in the vicinity but exceeds the theoretical percolation threshold on a fcc lattice with nearest-neighbor (NN) interaction of $\simeq 0.2$~\cite{vanderMarck1997}. 
and hence presents access to study the evolution of magnetic order in a material with fixed dilution fraction where the spin size, magnetic anisotropy, and orbital configuration are changed. In addition, due to the higher energy and larger spatial extent of the Se 4$p$ orbitals compared to the S 3$p$ orbitals, the substitution of S by Se modifies the $M$–$Ch$ bond length and is supposed to reduce the charge transfer energy~\cite{hu2015}, both of which being further parameters potentially controlling magnetism in \LiMChO . 

The overwhelming majority of studies on Li-rich antiperovskites have focused on their potential for applications: They are discussed as solid electrolyte and solid state battery materials due to their superionic conductivity and potential multi-electron transfer~\cite{lu2018,zhao2012,fang2017}. In addition, a variety of studies was performed to investigate the feasibility of several members of the \LiMChO -family as cathode materials in lithium-ion batteries~\cite{lai2017,lai2018,Gorbunov2020,Gorbunov2024,Singer2023elucidating,Mohamed2023,dai2025}. Recently, we confirmed long-range antiferromagnetic (AFM) order as well as preceding short range order in \LiFeSO\ and \LiFeSeO\ by means of magnetic susceptibility, NMR and Mössbauer studies~\cite{Seewald2025}.
 
In this work, we investigate the magnetic properties of \LiMChO\ compounds ($M$ = Fe, Co, Mn; $Ch$ = Se, S) by means of magnetic susceptibility, isothermal magnetization, and X-band electron spin resonance (ESR) measurements. %Typically, all Li-rich antiperovskites reported in the literature exhibit tiny amounts of ferromagnetic (FM) impurities which is likely associated with their metastable nature and the presence of stable competing phases.~\cite{lu2018,lai2018,Gorbunov2020,Mohamed2023mechanochemical,Mohamed2023,Singer2023elucidating,Singer2024,Gorbunov2021,Gorbunov2024,dai2025,graessler2025} 
The intrinsic magnetic susceptibilities are  deduced from the high-field magnetization $M$, obtained both isothermally, i.e., from $M$ vs. $B$ curves, and in static magnetic field when sweeping the temperature. Our data exclude a Curie- or Curie-Weiss-like response typically observed in materials featuring localized magnetic moments. We find an only very weak temperature-dependence of the intrinsic magnetic susceptibility $\chi_0$ in all materials. The observed values of $\chi_0$ are high when compared to typical Pauli-magnets. We suggest that the magnetic response originates from a diluted localized magnetic sublattice but also discuss itinerant scenarios and the possibility of an associated electronic instability. All materials under study show long-range AFM order. The transition temperatures increase following the sequence of Mn, Fe, Co, while except for $M$ = Co, the chalcogenide has no sizable effect on $T_{\rm{N}}$. We conclude a significant magnetic coupling energy ($>300$~K). The reduction of \tn\ due to dilution of the magnetic subsystem is strongly affected by the magnetic anisotropy of the transition metals in \LiMChO .

%%%%%%%%%%%%%%%%%%%%%
\section{Experimental Methods} \label{methods}
%%%%%%%%%%%%%%%%%%%%%

\begin{figure}%[tb]
\begin{center}
 \includegraphics[width=\columnwidth,clip]{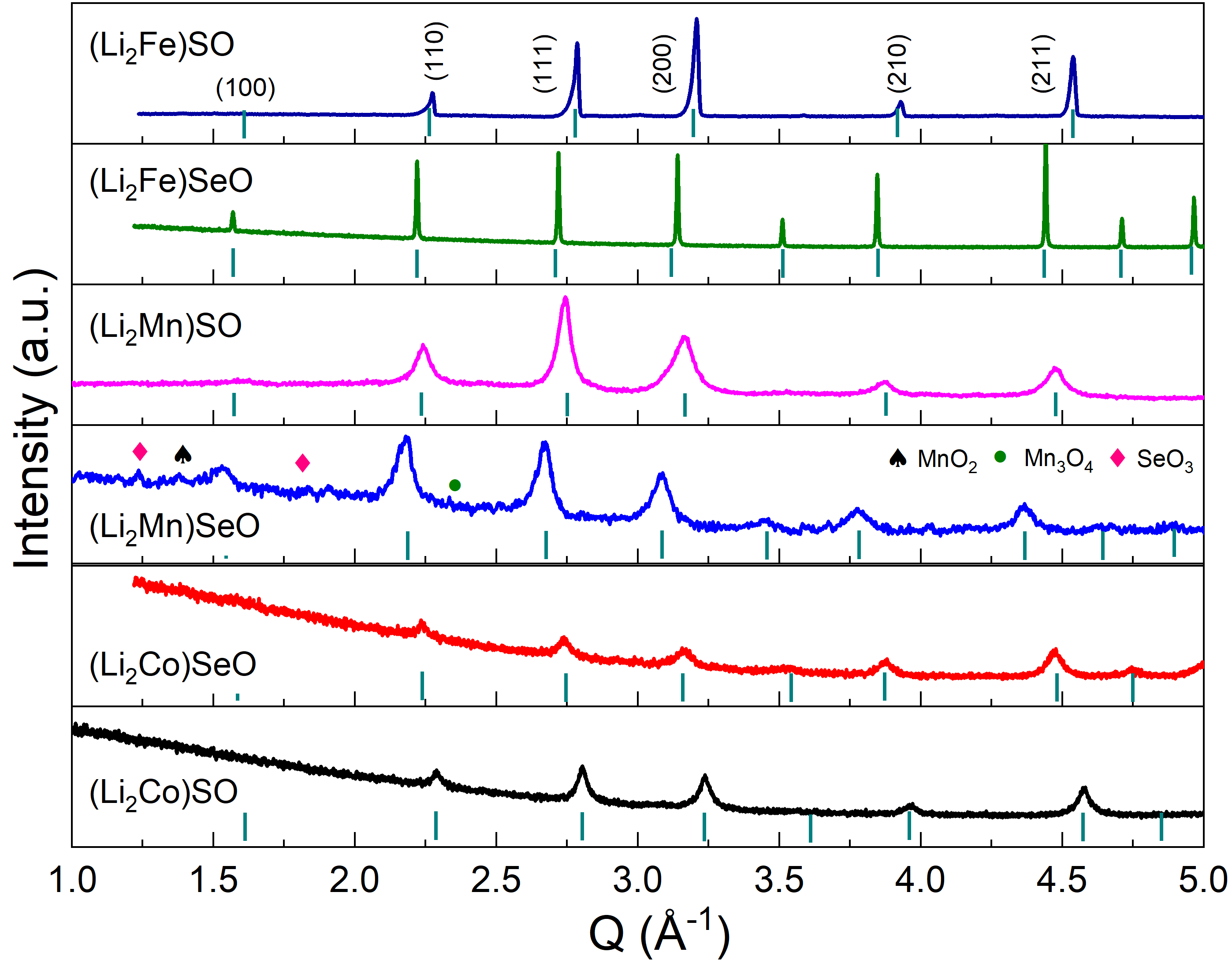}
 \caption{Powder X-ray diffraction patterns of the materials under study. Dark cyan lines in each plot mark the Bragg positions of the expected main phases (ICSD files for all materials under study: \LiFeSO: 253936; \LiFeSeO: 253937; \LiMnSO: 434436; \LiMnSeO: 434437; \LiCoSO: 434438; \LiCoSeO: 434439)~\cite{lai2017,lai2018}. {\rev The scattering vector $Q$ is used as the abscissa to compare data obtained using different radiation sources (see the text). Crystalline} impurity phases are detected in \LiMnSeO\ as shown by the markers. The data of \LiFeSO\ and \LiFeSeO\ are reprinted from Ref.~\cite{Mohamed2023,Singer2023elucidating}.}
\label{fig:xrd}
\end{center}
\end{figure}

The materials \LiMChO\ with $M$ = Fe, Co, Mn and $Ch$ = S, Se were synthesized as reported previously for $M$ = Fe and heat treated in different ways in order to alter various impurity phases. Tiny amounts{\rev \footnote{\rev For a detailed quantitative discussion see the Appendix.}} of impurity inevitably show up in this class of materials which is likely associated with their metastable nature and the presence of stable competing phases.~\cite{lai2018,Gorbunov2020,Mohamed2023mechanochemical,Mohamed2023,Singer2023elucidating,Singer2024,Seewald2025,Gorbunov2021,Gorbunov2024,dai2025,graessler2025}
%~\cite{Mohamed2023mechanochemical,Singer2023elucidating,Singer2024,Mohamed2023}. 
%Typically, all Li-rich antiperovskites reported in the literature exhibit tiny amounts of ferromagnetic (FM) impurities which is likely associated with their metastable nature and the presence of stable competing phases.~\cite{lu2018,lai2018,Gorbunov2020,Mohamed2023mechanochemical,Mohamed2023,Singer2023elucidating,Singer2024,Gorbunov2021,Gorbunov2024,dai2025,graessler2025}
The yet unreported synthesis of \LiMChO\ with $M$ = Mn, Co by means of a high-energy ball milling process is described in the %Supplement Material (SM)~\cite{SM}. 
Appendix. The prepared materials were analyzed by powder X-ray diffraction (pXRD) (STOE STADI P diffractometer) using the Debye–Scherrer method with Mo K$_{\alpha 1}$ radiation source ($\lambda = 0.70926$~\AA ) for the Mn-based samples or Co K$_{\alpha 1}$ radiation source ($\lambda = 1.7902$~\AA ) for the Co-based samples and a Mythen 1K detector (Dectris). The samples were filled into glass capillaries inside the glove box and melt-sealed to prevent any air exposure during XRD investigations. The corresponding pXRD patterns are shown in Fig.~\ref{fig:xrd}.
%the SM~\cite{SM}. 
Elemental analysis by inductively coupled plasma-optical emission spectroscopy (ICP-OES) (iCAP 6500 Duo View, Thermo Fisher Scientific GmbH) was used to estimate the stoichiometry of the produced compositions in terms of molar ratios of the elements{\rev $^1$}. Scanning electron microscopy (Nova-NanoSEM 200) coupled with energy dispersive spectroscopy (EDS Genesis with 15~kV accelerating voltage) was used to evaluate the morphology and composition of the studied compounds.

The successful synthesis of the stoichiometric materials is demonstrated by the pXRD patterns (Fig.~\ref{fig:xrd}) and by the ICP-OES results (Table.~\ref{tab:ICPOES}) of the as-ball-milled samples. The diffraction patterns exhibit broad but well-defined Bragg peaks that match the reported ICSD/CSD reference patterns for the corresponding compounds~\cite{lai2017,lai2018}. %\hl{The significant peak broadening indicates small crystallite sizes, consistent with high-energy mechanochemical synthesis. 
Minor crystalline secondary phases were only detected for \LiMnSeO . The detailed discussion of tiny impurity phases below the pXRD detection limit can be found in the Appendix. %SM~\cite{SM}. 
The ICP-OES data show that the compositions are in good agreement with the nominal stoichiometries.

Magnetic measurements were performed on powder samples using a MPMS3 SQUID (Superconducting Quantum Interference Device) magnetometer (Quantum Design). Prior to the measurements, the samples have been stored and mounted to sealed sample containers in an argon-filled glovebox. The static magnetic susceptibility $\chi =M/B$ has been obtained upon increasing the temperature at 1~T by using field-cooled (FC) and zero-field-cooled (ZFC) protocols where the sample was cooled either in the external measurement field or the field was applied after cooling to the lowest temperature. {\rev Measurements have been restricted to $<330$~K; the upper temperature limits for the materials were chosen to avoid thermal hysteresis effects which we attribute to thermal degradation of the metastable materials.} Isothermal magnetization $M(B)$ has been measured at $T=1.8$~K in the field range -7~T~$\leq B\leq$~+7~T. {\rk In addition, continuous-wave electron spin resonance (ESR) measurements at a fixed microwave frequency of 9.56~GHz and in magnetic fields up to 0.9~T were performed, at room temperature, using a commercial X-band spectrometer (EMX from Bruker). The obtained ESR spectrum was fitted using Lorentzian and Dysonian line shapes including the contribution from the resonance at negative magnetic field.~\cite{Dengler2010}}

%%%%%%%%%%%%%%%%%%%%%%%%%%%%%%%%%%
\section{Results}
%%%%%%%%%%%%%%%%%%%%%%%%%%%%%%%%%%

\subsection{Static magnetic susceptibility and Fisher's specific heat}

\begin{figure} [tb] 
    \includegraphics[width = \columnwidth]{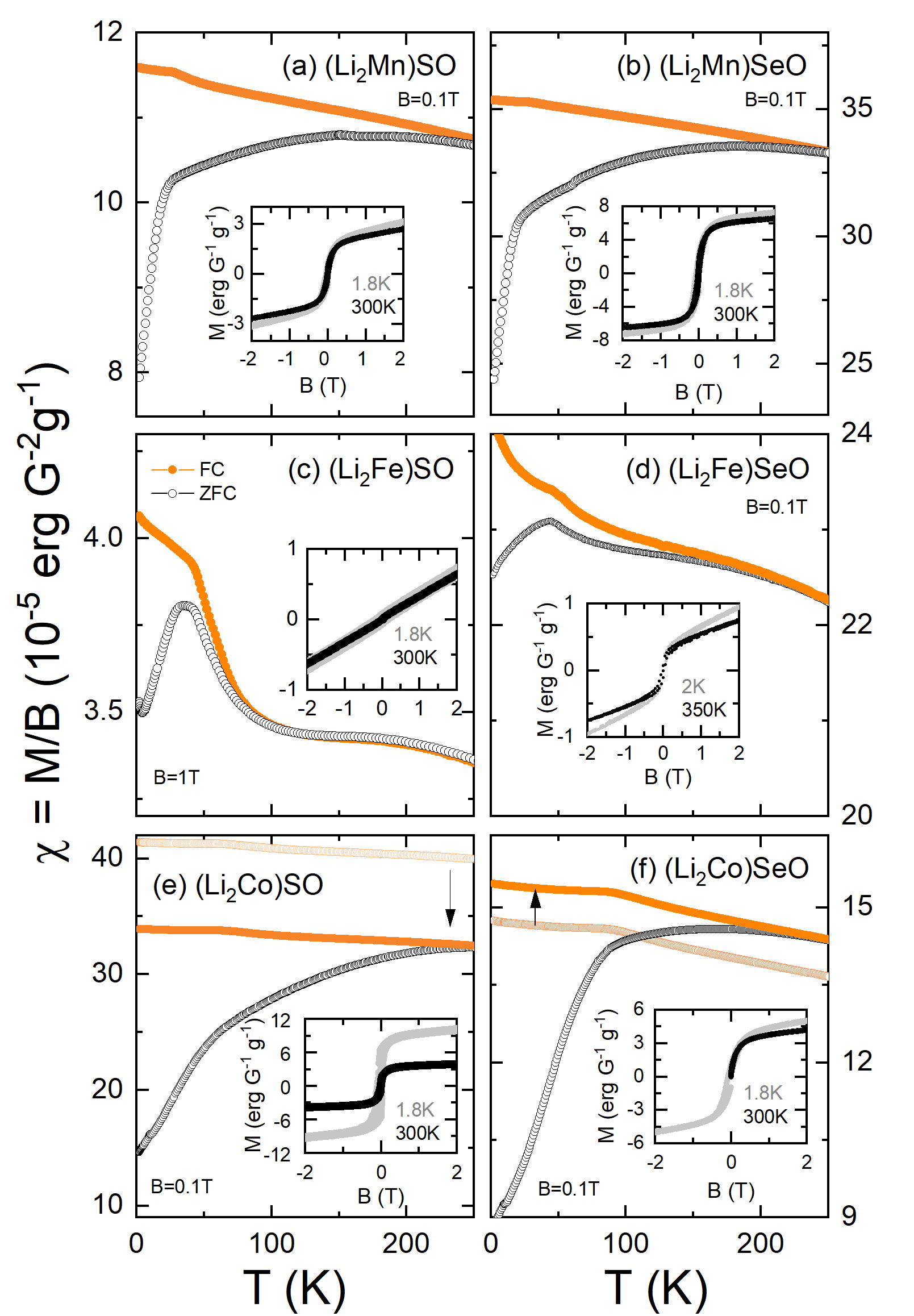}
    \caption{Static magnetic susceptibility ($\chi = M/B$) of \LiMChO\ with $M$ = Mn, Fe, Co and $Ch$ = S, Se. Both FC (filled orange markers) and ZFC (open black markers) protocols have been applied. In (e,f), FC data have been shifted to compensate for a hysteresis effect of a tiny FM impurity phase. Insets: Corresponding isothermal magnetization at 2~K and 300~K. 
    }\label{SQUIDall}
\end{figure}

The static magnetic susceptibility $\chi(T) = M(T)/B$ obtained both in ZFC ($\chi_{\rm ZFC}$) and FC ($\chi_{\rm FC}$) protocols as well as the isothermal magnetization $M(B)$ of \LiMChO\ with $M$ = Mn, Fe, Co and $Ch$ = S, Se are shown in Fig.~\ref{SQUIDall} and represent the magnetic response of Li-rich antiperovskites upon the variation of the $M$- and $Ch$-ions. The main features are as following: (1) In the whole temperature regime under study, \chifc\ decreases upon heating but displays only a rather small or even constant temperature dependence, except for a step- or kink-like behavior at low temperatures. In particular, we do not observe a general Curie-Weiss-like behavior in the accessible temperature range which would clearly indicate the presence of localized magnetic moments. Notably, the seemingly small Curie-like contribution in \lifeseo\ at lowest temperatures is not present in the ZFC data which excludes attributing it to a simple Curie- or Curie-Weiss-like response. The absence of a mean-field-like response is further illustrated by the double-logarithmic plot of the temperature dependence of $\chi^{-1}$ (see Fig.~\ref{powerlaw} in the Appendix). %SM~\cite{SM}). 
We also note that our data do not show any activated behavior in the magnetic susceptibility which excludes the scenario of a narrow-bandgap semiconductor. (2) For all materials, we find that ZFC/FC splitting extends either to room temperature or to about 100~K (\lifeso ). In the vicinity of the anomalies in \chifc , the FC/ZFC splitting considerably increases which yields a clear maximum in \chizfc\ in (Li$_2$Fe)$Ch$O, at around 50~K, while this feature is replaced with a kink and a subsequent smooth increasing, giving rise to an arc-like hump in (Li$_2$Mn)$Ch$O and (Li$_2$Co)$Ch$O. (3) Isothermal magnetization confirms small FM impurities (see the insets of Fig.~\ref{SQUIDall}). The nature and content of potential impurities are discussed in detail in the Appendix and range between nearly vanishing impurity fractions $\ll 0.1$~w\% up to fractions of $\simeq 5$~w\% (see Table~\ref{tab:impurities}).

\begin{figure} [hbt] 
    \includegraphics[width = \columnwidth]{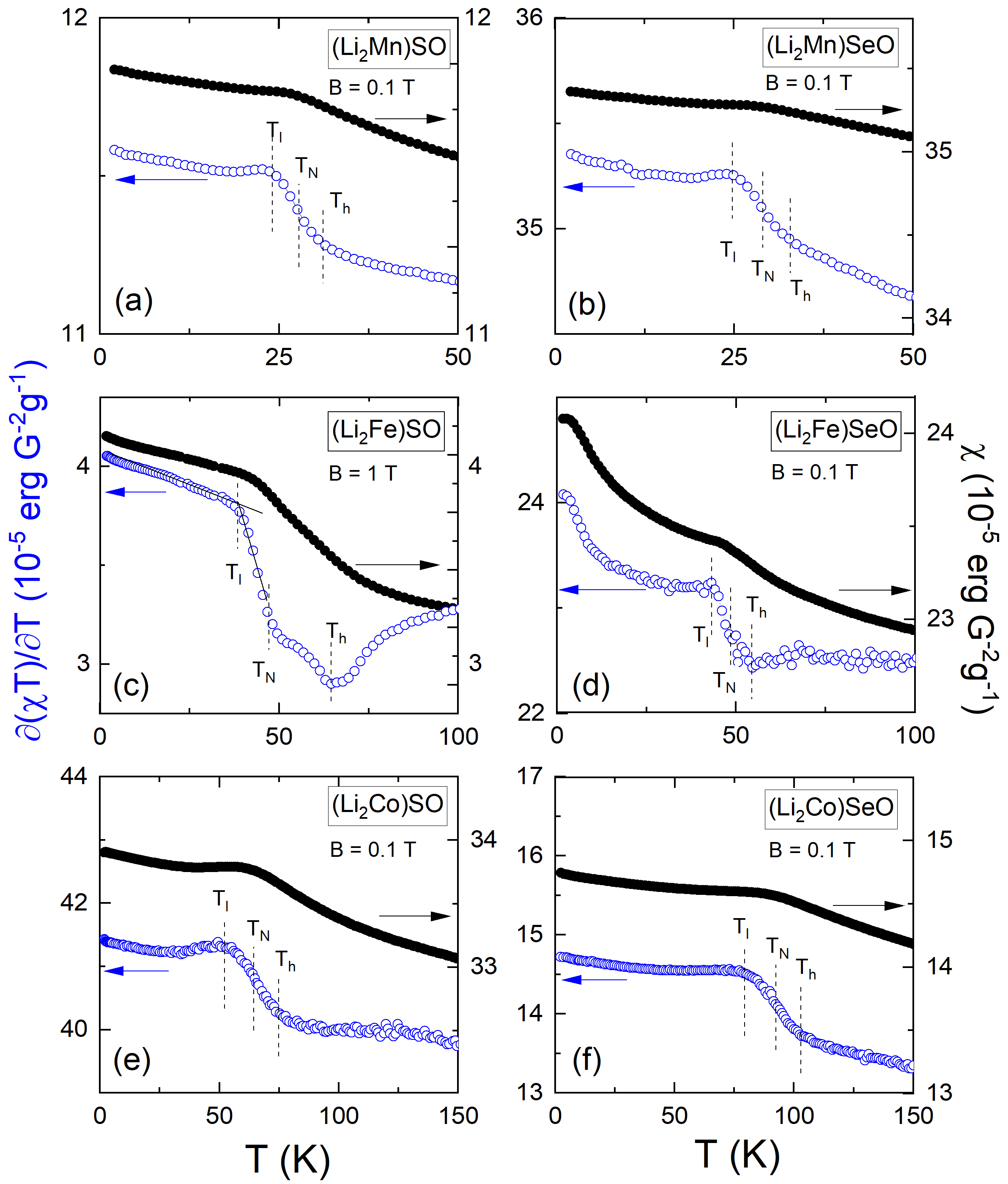}
    \caption{Fisher's specific heat $\partial (\chi T)/\partial T$ (blue open markers) as derived from the FC static magnetic susceptibility (black filled markers) in Fig.~\ref{SQUIDall}. The center of the step marked by $T_{\rm N}$ signals the evolution of long-range magnetic order while \tright\ and \tleft\ are the temperatures where the steps begin and end.} %\hl{reorder subfigures consistent with Fig2}} %\hl{I updated the values and positions}}
        \label{cpmag}
\end{figure}

\begin{figure} [tb] 
    \includegraphics[width = 1\columnwidth]{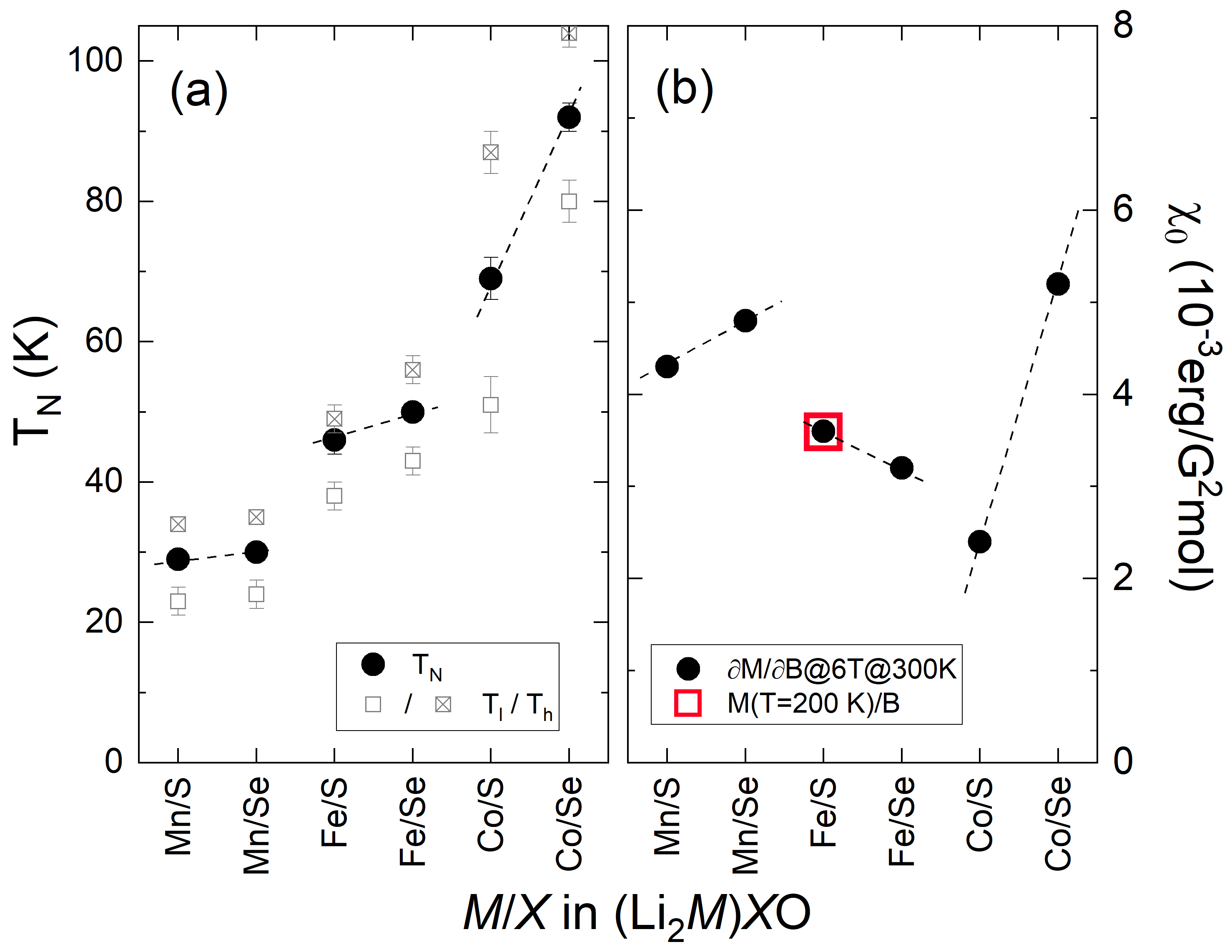}
    \caption{(a) Magnetic ordering temperature \tn\ and (b) magnetic susceptibility $\chi$ of (Li$_2M$)$Ch$O ($M$ = Mn, Fe, or Co; $Ch$ = Se or S). \tleft\ and \tright\ indicate the low- and high-temperature bounds of the jumps in Fisher's specific heat as illustrated in Fig.~\ref{cpmag}. $\chi$ has been determined as $\partial M(B)/\partial B$ at $B\simeq 6\,{\rm T}, T =300\,{\rm K}$ and (only for \LiFeSO\ as described in the text) from $M(T)/B$ (at $B=1\,{\rm T})$. Lines are guide to the eye.}
    \label{fig:tn}
\end{figure}

Calculating Fisher's specific heat $\partial (\chi_{\rm FC} T)/\partial T$ allows us to estimate the temperature dependence of the  magnetic entropy changes (see Fig.~\ref{cpmag})~\cite{fisher1962relation}. The results show a step-like feature in the magnetic specific heat of all studied materials, which is indicative of the onset of long-range magnetic order. For Li$_2$Fe$Ch$O, local probe studies, i.e., NMR and Mössbauer spectroscopy, recently confirmed the assignment of the observed step in $\partial (\chi_{\rm FC} T)/\partial T$ to the onset of long-range AFM order~\cite{Seewald2025}. For completeness, we characterize the step-like features by their onset temperatures, \tright , a tiny maximum at the low-temperature edge of the step, \tleft , and the step-center which we attribute to the actual ordering temperature, \tn . The obtained values are summarized in Table~\ref{tab:tn} and are displayed in Fig.~\ref{fig:tn}.

\begin{table}[tb]
    \centering
    \begin{tabular}{l|cccc}%
    %\begin{tabular}{p{0.1\textwidth}p{0.1\textwidth}p{0.1\textwidth}p{0.1\textwidth}p{0.1\textwidth}}
    \hline\hline
         &~\tn~ & ~\tleft~ & ~\tright~ & ~$\chi_0$~ \\
         &(K)&(K)& (K) &$10^{-3}\frac{\rm erg}{\rm G^2mol}$\\
         \hline
       \LiMnSO & ~28(2)~ & ~23(2)~ & ~32(2)~ & 4.5(1) \\
       \LiMnSeO~ & ~29(2)~ & 25(2) & 32(2) & 4.7(2)\\
       \LiFeSO & ~46(2)~ & 38(2) & 64(2) & 3.6(1)\\
       \LiFeSeO & ~50(2)~ & 43(3) & 56(2) & 3.2(1)\\
       \LiCoSO & ~67(3)~ & 55(4) & 77(2) & 2.37(5)\\
       \LiCoSeO & ~92(2)~ & 83(3) & 101(2) & 5.2(1)\\
    \hline\hline
    \end{tabular}
    \caption{Characteristic temperatures of the magnetic specific heat anomalies and high-field paramagnetic suceptibilities
    $\chi_0=\partial M/\partial B$ at 6~T. %Error bars in $\chi_0$ mainly reflect the uncertainty in the mass determination and sample holder signal.~\footnote{Additional uncertainties of $\chi_0$ from impurity phases are small in comparison ($\lesssim 10^{-4}$~erg/(G$^2$mol)) and are discussed in detail in the SM~\cite{SM}.} 
    In addition to the long-range ordering temperature \tn, which is determined as the center of the step-like increase in Fisher's specific heat, we also list \tright\ and \tleft\ the step begins and ends (see Fig.~\ref{cpmag}).}
    \label{tab:tn}
\end{table}

\subsection{Paramagnetic susceptibility}

All \LiMChO\ compounds presented in the literature feature magnetic impurity phases~\cite{lai2018,Gorbunov2020,Mohamed2023mechanochemical,Mohamed2023,Singer2023elucidating,Singer2024,Gorbunov2021,Gorbunov2024,Seewald2025,dai2025,graessler2025}. This is most likely due to the metastability of the materials as suggested by their partial decomposition into binary phases at intermediate temperatures and supported by DFT studies~\cite{lai2017,lu2018,Zhu2022}. Both binary oxides and chalcogenides are stable competing phases~\cite{wang2020}. While this complicates, or even entirely prevents, the quantitative determination of the comparatively small paramagnetic response from \textit{low-field} magnetization measurements, the vanishing magnetic susceptibility of ferromagnets at sufficiently \textit{high} magnetic fields enables us to determine the intrinsic paramagnetism of the main \LiMChO\ phases\footnote{{\rev A quantitative estimate of the in comparison vanishing magnetic susceptibility of the impurity phases in sufficiently high magnetic fields can be found in the Appendix.}}. We hence have determined the high-field magnetic susceptibilities $\chi_{0}=\partial M(B\approx 6\,{\rm T})/\partial B$ from the isothermal magnetization data obtained at room temperature. The resulting susceptibilities are displayed in Table~\ref{tab:tn}.\footnote{Error bars in $\chi_0$ mainly reflect the uncertainty in the mass determination and due to the sample holder signal. Additional uncertainties of $\chi_0$ from impurity phases are small in comparison ($\lesssim 10^{-4}$~erg/(G$^2$mol); see the Appendix).} %and are discussed in detail in the SM~\cite{SM}.} 
The particular low amount of FM impurities in \lifeso\ additionally enables us to compare the static and differential magnetic susceptibilities for this material. To be specific, we find a tiny FM response at 300~K of $M_{\rm imp}=3\times 10^{-2}$~erg/(Gg)\ and subtract it from the measured data.\footnote{We attribute the FM response to a Fe$_x$S impurity phase  
with $x \leq 1$ with phase fraction of a few (1-3) per mille which is invisible in conventional characterization techniques.} 
%Its typical saturation magnetization of $10-20$~\ergGg\ ~\cite{volk2016,kind2013} 
After correcting for the small FM signal ($M_{\rm imp}/M_{\rm exp}\simeq 0.1$), the static magnetic susceptibility $(M-M_{\rm imp})/B$, at $T=200$~K, perfectly agrees to $\chi_{0}=\partial M(B\approx 6\,{\rm T})/\partial B$ (see the red marker in Fig.~\ref{fig:tn}). Furthermore, this agreement confirms that the high field susceptibilities for all six samples are reliably extracted from the isothermal magnetization (see Tab.~\ref{tab:tn}). %Notably, the systematic error bar exhibited in Tab.~\ref{tab:tn} excludes the specific contribution of impurity phase which will be discussed in the SM).

\section{Discussion}

\subsubsection*{Magnetic response well above $T_{\rm N}$}

\begin{figure} [tb] 
    \includegraphics[width = \columnwidth]{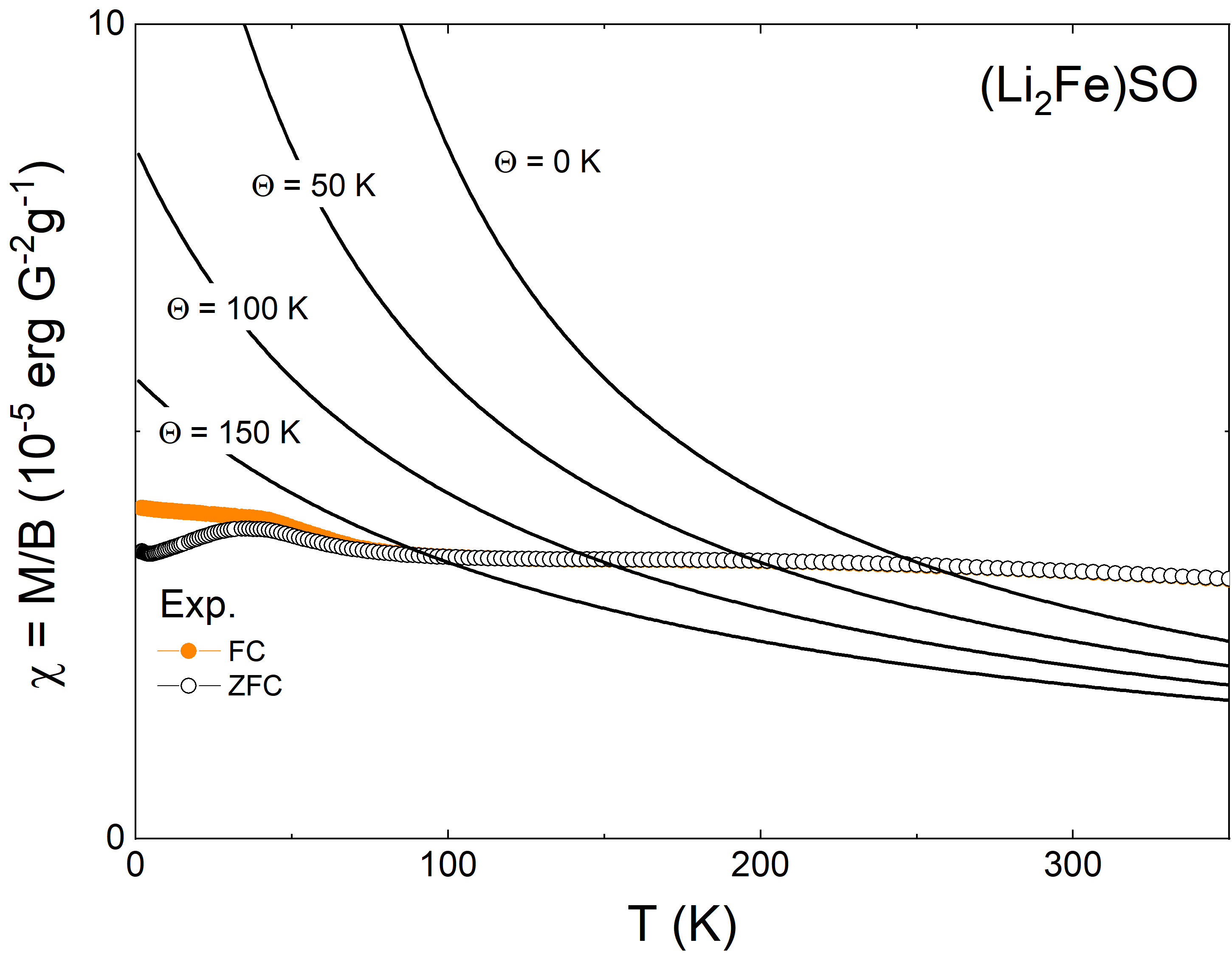}
    \caption{Experimental magnetic susceptibility of \LiFeSO\ (data points) from Fig.~\ref{SQUIDall}c and calculated magnetic response for a system of localized moments $S=2$ on the Fe-positions according to the Curie-Weiss law ($\chi=C(T+\Theta)$) with the Curie constant $C$ and varying Weiss temperatures $\Theta$ (lines). $\Theta=0$ reflects non-interacting moments while $\Theta>0$ accounts for effective antiferromagnetic coupling. %The $g$-factor $g=2$ was used for the estimates. 
    Note that, for $T<\Theta$, Curie-Weiss behavior is not expected in the mean-field model, which is not reflected by the shown Curie-Weiss curves. 
    }
    \label{Comparison}
\end{figure}

The absence of a Curie- or Curie-Weiss-like response (see also Fig.~\ref{Comparison}) in the magnetic susceptibility, that would clearly signal the presence of localized magnetic moments, challenges the prevalent model of \LiMChO\ as an insulator with a band gap of $\gtrsim 2$~eV as based on previously reported numerical studies~\cite{lu2018,Gao2023,ni2024voltage,Gorbunov2024correlation}. Specifically, large localized moments of 4~\mb /Fe and a band gap $E_{\rm g}$ of $\sim 2.7$~eV are predicted for \LiFeSO~\cite{lu2018}. This has been confirmed by various numerical studies where band gaps $\gtrsim 2$~eV  %(for \LiMnSO\ and \LiCoSO~\cite{lu2018}, \LiFeSO\ and \LiMnSO~\cite{ni2024voltage}) 
are found.\cite{lu2018,ni2024voltage} $E_{\rm g}$ of 1.5 to 2~eV is also predicted for all (Li$_2M$)SeO studied at hand~\cite{Gorbunov2024correlation}. We note that Ref.~\cite{zarhri2024ab}, in contrast, suggests metallic behavior in \LiFeSO. The rather Pauli-like magnetic susceptibility found for all materials in our study, hence, might suggest the presence of conduction electrons in all \LiMChO\ under study, i.e., a finite density of state (DOS) at the Fermi energy, but -- as will be explained below -- is also consistent with the notion of localized spins. In the following, we will discuss the scenarios of diluted localized magnetic moments residing on the X-positions of the antiperovskite \abx\ structure as well as itinerant electron scenarios.

%\subsubsection{Localized spin scenarios}

\begin{table*}
    \centering
    \begin{tabular}{l|c|c|c|c|c|c|c|c}
    \hline\hline
     & $d_{\rm X-X}$ (\AA) & $d_{M{\rm -O}}$ (\AA) & $\angle$($M$-O-$M$) (\degree)& $\angle$($M$-$Ch$-$M$) (\degree) & $t$ & $S$ & \tn\ (K) & $\frac{3 T_{\rm N}}{S(S+1)}$ (K)\\
    \hline
    \LiMnSO   & 2.80700(11) & 1.98485(8)  & $90^\circ$ & $60^\circ$ & 0.8496 & 5/2 &29&$\sim10$\\
    \LiMnSeO  & 2.87143(8)  & 2.03041(6)  & $90^\circ$ & $60^\circ$ & 0.895 & 5/2 &30& $\sim10$\\
    \LiFeSO  & 2.76758(0)  & 1.95697(0)  & $90^\circ$ & $60^\circ$ & 0.8507 & 2  &46&$\sim23$\\
    \LiFeSeO  & 2.8302(3)   & 2.00125(16) & $90^\circ$ & $60^\circ$ & 0.8964 & 2  &47&$\sim24$\\
    \LiCoSO   & 2.74990(10) & 1.94447(8)  & $90^\circ$ & $60^\circ$ & 0.8515 & 3/2  &69&$\sim55$\\
    \LiCoSeO  & 2.80806(6)  & 1.98559(5)  & $90^\circ$ & $60^\circ$ & 0.8974 & 3/2 &92&$\sim74$ \\
    \hline\hline
\end{tabular}
\caption{Distances $d_{\rm X-X}$ between the crystallographic $X$-positions in the \abx\ structure (X = $M$, Li) and between $M$ and O-sites ($d_{M{\rm -O}}$) as well as associated bond angles and the Goldschmidt tolerance factors $t$. $t$ is defined as $t=\frac{r_{Ch}+r_{Li_2M}}{\sqrt{2}(r_{O}+r_{Li_2M})}$ using the average radius of $Li_2M$~\cite{lai2017, lai2018}. $S$ is the expected spin state of $M^{2+}$ and \tn\ is the N\'eel temperature from table~\ref{tab:tn}. The last column is proportional to the mean-field estimate of the effective magnetic coupling ($ zJ_{\rm eff}$) using $k_{\rm B}T_{\rm N}\sim z |J_{\rm eff}| S(S+1)/3$, with $k_{\rm B}$ being the Boltzmann constant, $z$ the number of nearest neighbors, and $S$ the spin quantum number in the high-spin state.\footnote{Note, that this estimate is based on the Heisenberg model which might be valid only to a limited extent for $M$ = Fe, Co.} %{\lb indeed, its 3 kB TN = zJS(S+1), see e.g. Blundell (5.17) with this formular for ferromagnets. So the last column of the table should be 3TN/S(S+1).   } \hl{recalculate last column} %{\jy Jieyuan:  The eight nearest neighborships involve a 60\degree antiferromagnetic exchange via two (S, Se) ions. This may be saying there is two exchange path between two magnetic ions which are in neighbor. i.e. M-Ch1-M; M-Ch2-M. Ch1, Ch2 are two corner ions (S or Se) being neighbor of Magnetic ion. So the path is 60\degree M-Ch-M.}
}
\label{tab:structure}
\end{table*}

%Firstly, we note that we do not find activated behavior in the magnetic susceptibility which excludes the scenario of a narrow-bandgap semiconductor. 

The possible two main exchange paths between neighboring localized magnetic moments in \LiMChO\ are a $M$-$Ch$-$M$ path with 60\degree\ bond angles (which suggests its AFM nature) and a (presumingly weak FM) 90\degree\ exchange path via an oxygen ion (see Tab.~\ref{tab:structure}). The observed long-range AFM order suggests the dominance of the former exchange path. The scenario of localized moments and semiconducting behavior is not excluded by our observation of the only weakly temperature-dependent magnetic susceptibility and clear non-Curie-Weiss behavior: One possible explanation for such a behavior is the presence of differently sized, finite magnetic clusters. When the  magnetic interaction is sufficiently large, such clusters exhibit size-dependent excitation gaps and reduced magnetization. The size-dependent distribution of excitation gaps can in principle account for unexpected and weakly temperature-dependent magnetic susceptibility (see examples in Refs.~\cite{Vavilova2006,Garcia2015,Koo2016,reis2006,deisenhofer2006}). We also note that antiferromagnetic interaction of localized spins has also been found in itinerant systems, as e.g. observed in CaCu$_3$Ru$_4$O$_{10}$, where immersion of Cu$^{2+}$ ions in an itinerant band and associated Kondo physics are suggested to yield strongly reduced and weakly temperature-dependent spin susceptibility~\cite{Takegami2022}.

In the following, we further discuss this scenario of differently sized, finite magnetic clusters. A particular property of the Li-rich antiperovskites studied here is their strongly disordered nature arising from the shared position of the Li- and $M$-ions in the \abx\ structure (i.e., \LiMChO ). Recent X-ray pair distribution function (PDF) and Mössbauer studies suggest a nearly random occupation of the X-position~\cite{Coles2023,Deng2023,Seewald2025} by both metal species. For 1/3-occupation with magnetic centers, the probability to find isolated clusters (magnetic multimers) is hence directly related to the probability of finding $n$ nearest magnetic $M$-sites on the nearest-neighbor (NN) X-positions which is given by $f_n = \frac{8!}{(8-n)!n!} \times (\frac{1}{3})^n \times (\frac{2}{3})^{8-n}$.~\cite{Seewald2025}
%\begin{equation}
 %           f_n = \frac{8!}{(8-n)!n!} \times (\frac{1}{3})^n \times (\frac{2}{3})^{8-n}.
  %          \label{eq:probability}
%\end{equation}
This allows us to calculate the probabilities to find isolated clusters of paramagnetic $M$-ions of size $m$.\footnote{Here, cluster size means the number of magnetic sites in the cluster.} We find that 3.9~\% of all $M$-sites form magnetic monomers (i.e., $\simeq 1.3$~\% of all $X$-sites). For magnetic dimers, exactly one of the NN sites of $M$ is magnetic, too, which is realized in about 2~\% of the magnetic sites\footnote{$f_{\rm dim} = f_1 = 8 \times (\frac{1}{3}) \times (\frac{2}{3})^{12}.$}. For trimers, four configurations are possible (equilateral triangles (eqT) where all three magnetic ions are NN; $L$-shaped triangles (LT) where the center $M$-ion has two nearest $M$ neighbors which are second NN; linear trimer chains (linT); other triangles (othT)\footnote{The 'other triangles' refer to those where the angle opposite of the longest side is neither 60\degree\ nor 90\degree .}) The probability for trimers is hence $\sim 1.1$~\%.\footnote{$f_{\rm eqT} = 8\times(\frac{2}{2}) \times (\frac{1}{3})^{2} \times (\frac{2}{3})^{15}\simeq 0.2~\%;    f_{\rm LT} = 8\times(\frac{4}{2}) \times (\frac{1}{3})^{2} \times (\frac{2}{3})^{15}\simeq 0.4~\%;  f_{\rm linT} = 8\times(\frac{2}{2}) \times (\frac{1}{3})^{2} \times (\frac{2}{3})^{16}\simeq 0.12~\%;    f_{\rm othT} = 8\times(\frac{4}{2}) \times (\frac{1}{3})^{2} \times (\frac{2}{3})^{15}\simeq 0.4~\%.$
} Larger clusters become increasingly improbable. %{\jy comparing with the trimer case, the probability should reduce to $$(\frac{1}{3}) \times (\frac{2}{3})^3$$ roughly} 

The valence of the metal ions has been spectroscopically confirmed to be +2~\cite{mikhailova2018,Gorbunov2021}. 
While, in principle, for Co$^{2+}$ ($S=3/2, 1/2$) and Fe$^{2+}$ ($S=2,1$) several spin states may be possible\footnote{The low-spin state of Fe$^{2+}$ has been experimentally excluded~\cite{Seewald2025}.}, recent XPS data suggest high-spin configurations~\cite{mikhailova2018,Mohamed2023}. 
The superposition of the responses of differently-sized \textit{small} finite spin ensembles, i.e., the superposition of responses characterized by different excitation gaps can yield a weakly temperature dependent magnetic susceptibility.\cite{Garcia2015,reis2006,deisenhofer2006} The small probability of finding isolated small clusters with $n\geq 2$, however, strongly suppresses the experimental signature of the superimposed spin gaps (for illustration, see Fig.~\ref{fig:cluster} in the Appendix where the susceptibilities of finite spin chains have been added up according to their occurrence as calculated above). Depending on the local arrangement of the $M$-sites, and as illustrated by the considerations on the various trimer configurations exemplarily discussed above, geometric magnetic frustration may be relevant, too. In addition, the fact that no distinct Curie-tail is observed in the experimental data presented in Fig.~\ref{SQUIDall}
however implies that the expected monomers (and, more generally, odd-numbered spin-ensembles, such as trimers) do $not$ respond as isolated systems, thereby suggesting that next-nearest neighbor (NNN) interactions must be taken into account, too. %The presence of relevant NNN coupling has been also suggested by recent Mössbauer data~\cite{Seewald2025}. 
In conclusion, the presence of finite sized ensembles which are weakly-coupled via NNN interaction, frustration, and a disordered percolating network of isolated spins may hence qualitatively explain the observed temperature dependence of the magnetic susceptibility. This, however, requires comparably (to \tn ) large magnetic coupling since localized moments obey the Curie-Weiss behavior in the high-temperature limit. For \lifeso , our data would, thus, imply a strong effective magnetic NN coupling of $J > 300$~K.

\subsubsection*{Discussion of \tn }  

Long-range magnetic order is a common feature of (non-diluted) antiperovskite materials. A few FM and collinear AFM antiperovskites have been reported, while more than a dozen antiperovskites show non-collinear antiferromagnetism.~\cite{Singh2021,Deng2023a,Shamova2026} For example, Mn$_3$ZnN exhibits long-range AFM order below 183~K and an AFM-AFM transition at 140~K.~\cite{Deng2023a} In contrast to these materials where the X-sites are fully occupied by paramagnetic ions, the observation of a long-range magnetically ordered ground state may be less expected in the Li-rich materials where only 1/3 of the X-sites are paramagnetic.
The shared X-position by the Li- and $M$-ions yields significant dilution of the magnetic sublattice and strong disorder. Still, all \LiMChO\ discussed here display clear signatures of long-range AFM order at low temperatures. While the existence of long-range AFM order has been further confirmed by NMR and Mössbauer studies for \LiFeSO\ and \LiFeSeO~\cite{Seewald2025}, Fisher's specific heat allows us to follow the ordering temperature for the different compositions. Our data show that \tn\ is lowest for \LiMnChO , increases for \LiFeChO\ and is highest in \LiCoChO\ (cf. Fig.~\ref{fig:tn}). We also find that, except for $M$ = Co, the choice of the chalcogenide has no sizable effect on \tn . 

%In general, replacing S with Se is supposed to elevate the $M^{2+}$ energy levels due to the weak interaction between the higher energy level 4$p$ and 5$p$ chalcogenide states and the  3$d$ metal states~\cite{ni2024voltage}. 

We firstly discuss the very presence of long-range magnetic order which implies that the fraction of magnetic sites in all investigated materials overcomes the percolation limit. In the \abx\ antiperovskite structure, magnetic ions occupy 1/3 of the cubic face-center X-sites in \abx . They form a lattice whose magnetic percolation threshold has not been reported in the literature yet. It, however, features the same coordination number of nearest neighbors $Z_{\rm NN}=8$ and next nearest neighbors $Z_{\rm NNN}=6$ as the bcc lattice. As a first approximation we hence infer that the percolation threshold is close to that of the bcc lattice, which is $\simeq 0.246$~\cite{Xu2013}. For an arbitrary cubic lattice, the percolation threshold can be estimated by~\cite{balankin2019} 

\begin{equation}
    p_c = \frac{1}{1 + cQ \ln[Q^{0.5}(Z_\infty - 1)]},\label{eq:perc}
\end{equation}

where $Q$ is the connectivity index related to the dimension $d$ of the system by $Q=d-1$, $Z_\infty=8$ is the mean coordination number, and $c=1/\ln5$.~\cite{balankin2019} For \abx\ studied at hand, this yields an estimated percolation threshold of $p_c \simeq 0.26$. Both, the comparison with the bcc lattice as well as the numerical estimate using Eq.~\ref{eq:perc}, hence, place \LiMChO\ in the vicinity but slightly above the magnetic percolation limit which straightforwardly explains the observed evolution of long-range magnetic order. 

{\rev As shown in Table~\ref{tab:structure}, the variation of \tn\ neither corresponds to the changes in the distance or bond angles of the magnetic ions nor to the tolerance factor $t$. In particular, there are clear changes in the structure parameters when replacing S by Se which do not involve corresponding changes of \tn\ for $M$~=~Mn and Fe.} Considering the electronic structure, replacing S with Se is supposed to elevate the $M^{2+}$ energy levels due to the weak interaction between the high energetic 4$p$ and 5$p$ chalcogenide states and the 3$d$ metal states~\cite{ni2024voltage}. From a geometrical point of view, the larger Se$^{2+}$ ions yield a Goldschmidt tolerance factor closer to the ideal perovskite structure which might yield stronger AFM superexchange in \LiMSeO\ than in \LiMSO . Our data, however, suggest that a variation of the magnetic exchange coupling due to the structural changes in the series of materials under study is $not$ key for the observed changes in \tn . Instead, a relevant property may be magnetic anisotropy which is particularly small in Mn$^{2+}$ ions but typically strongly increases for Fe$^{2+}$ and Co$^{2+}$ due to the deviation from half-filling of the $t_{2g}$ orbitals. Accordingly, while \LiMnChO\ may be described as a (diluted) 3D Heisenberg magnet, the Ising character of the magnetic system progressively increases in \LiFeChO\ and \LiCoChO , with the latter exhibiting the most pronounced orbital contribution to the magnetic moment and, hence, the strongest Ising-character. 

\begin{figure} [tb] 
    \includegraphics[width = \columnwidth]{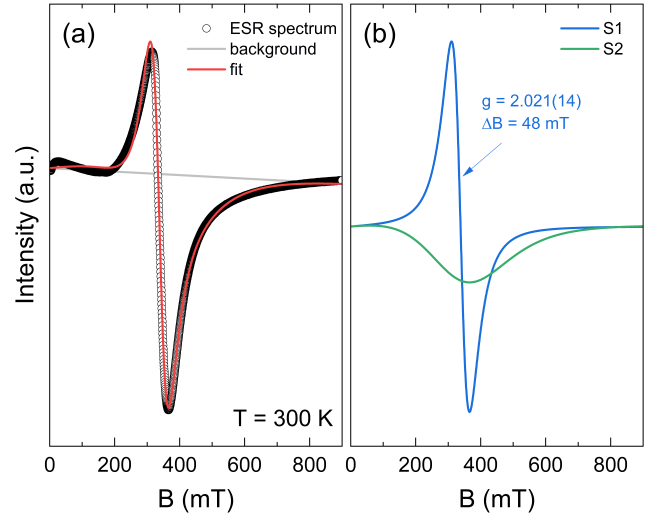}
    \caption{{\rk(a) ESR spectrum of \LiMnSO~at 300~K. The  central signal has been fitted (see (b)) by a pure Lorentzian line shape (signal S1), while a Dysonian line shape (signal S2) accounts for the broad variation of the signal background.}  }
    \label{fig:xepr}
\end{figure}

{\rk The presence of rather isotropic magnetic localized moments in \LiMnSO\ is confirmed by the ESR signal in Fig.~\ref{fig:xepr}. To be specific, the ESR signal is dominated by a Lorentzian-shaped line characterized by the $g$ factor $g=2.021(14)$. This is in the upper regime expected for Mn$^{2+}$ ions with $S=5/2$ and vanishing orbital momentum.~\cite{Krystek2006} The observed linewidth of $\Delta B=48$~mT is in the regime of what is observed in MnO ($\Delta B\approx 60$~mT (extrapolated from 200~K)~\cite{Golosovsky2005}), MnTiO$_3$ ($\Delta B\simeq 24$~mT~\cite{Gries2022}), LiMnPO$_4$ ($\Delta B\simeq 30$~mT~\cite{Wizent2009}), all of which at room temperature being in their paramagnetic regime. In addition to the clearly Lorentzian-shaped line, the spectrum exhibits an additional rather dispersive component which is described by an Dysonian line-shape (see Fig.~\ref{fig:xepr}b). One may speculate that it is associated with $\simeq 1$~\% mesoscopic (but not superparamagnetic) metallic iron inclusions in the material resulting from the synthesis process (see Table~\ref{tab:impurities} in the Appendix). Note, that we did not see X-band ESR resonance features in \LiFeChO\ (data not shown). The absence of both signals in the Fe-containing materials can be attributed to the strong single-ion anisotropy of Fe$^{2+}$, rapid spin–lattice relaxation, and the absence of metallic iron inclusions. The presence of the clear ESR signal in \LiMnSO~implies the presence of localized spin and suggests its non-metallic character.}

The dilution of the magnetic lattice prevents \tn\ to be taken as a direct measure of magnetic coupling between the magnetic moments. For example, (site-)diluting the 3D-Ising AFM Fe$_2$F and \mnf\ with diamagnetic Zn yields strong suppression of long-range magnetic order as well as softening and broadening of the associated anomalies in the magnetic susceptibility and the specific heat.~\cite{Sousa2010,Belanger1980,Bakshi1982}. In the example of \fezn\ with $x\simeq2/3$, which is  above the percolation threshold of this system, long-range AFM order %is strongly suppressed as compared to the undiluted material and 
evolves at only $\simeq 0.2\times T_{\rm N}(x=0)$.~\cite{Sousa2010} In addition, even in less diluted systems ($x=0.31$), \tn\ is significantly reduced and spin-glass-like behavior has been found at temperatures well below \tn ($x=0.31$) as evidenced by bifurcation in the FC/ZFC magnetic susceptibility~\cite{Montenegro1988}. A similar trend of reduction of \tn\ is seen in the fcc AFM system \comg\ where $T_{\rm N}(x=1/3)/T_{\rm N}\lesssim 0.1$~\cite{Kannan1986}. In the example of Fe$_y$Al$_{1-y}$, the long-range ordering temperature is reduced by an order of magnitude at about 60\% filling (dilution $y\simeq 0.6$) as compared to non-diluted materials~\cite{santos2021,yelsukov1992}. Typically, dilution of the magnetic subsystems initially leads a quasi-linear suppression of the magnetic ordering temperatures while suppression beyond linearity is found in the vicinity of the percolation threshold $p_c$~\cite{Sousa2010,Kannan1986,yelsukov1992}. We must, therefore, conclude that \tn\ of the hypothetical undiluted versions of the materials, i.e., Fe$_3Ch$O, Mn$_3Ch$O and Co$_3Ch$O, would be much higher than observed for the 2/3-diluted materials studied at hand. None of them, however, forms a stable structure so that we indirectly infer from the literature data on the above-mentioned diluted systems that $T_{\rm N}^{\rm Li-free}> 300$~K, i.e., more than an order of magnitude larger than \tn ($x=0.33$) found at hand. The expected high ordering temperatures of the non-diluted materials are consistent with recent findings on cation-deficient $\beta$-Fe$_2$SeO.~\cite{Shamova2026} Despite clear structural discrepancies~\cite{Krivovichev2024}, it is meaningful to compare the onset of long-range AFM order at $T_{\rm N} = 104$~K in 2/3-filled $\beta$-Fe$_2$SeO with significantly smaller $T_{\rm N}^{\rm (Li_2Fe)SeO} \simeq 47$~K where, e.g., the distances between the magnetic Fe$^{2+}$ species ($d_{\rm Fe-Fe}\simeq 2.93-2.97$~\AA) may suggest smaller magnetic coupling. 

While \tn\ does not directly reflect the NN magnetic coupling, it still may be useful to exploit the relation ${\lb 3}k_{\rm B}T_{\rm N}\sim z |J_{\rm eff}| S(S+1)$ obtained from the mean-field approximation of the Heisenberg model. Here, $k_{\rm B}$ is the Boltzmann constant, $z$ the number of nearest neighbors (NN), and $S$ the spin quantum number. The ratio ${\lb 3}T_{\rm N}/(S(S+1))$ given in table~\ref{tab:structure}, hence, reflects both the effective magnetic coupling provided by potentially competing exchange paths and the effective number of NN (i.e., $z\times |J_{\rm eff}|$). As mentioned above, the typical reduction in Mn-based diluted magnetic systems~\cite{Belanger1980,Bakshi1982} suggests that the hypothetical ordering temperature in Mn$_3Ch$O is $>300$~K. This would imply $|zJ_{\rm eff}|>100$~K. Such a strong magnetic coupling is consistent with our data which do not show CW behavior even at 300~K. {\rk The presence of strong magnetic coupling is further corroborated by the width of the room-temperature ESR signal presented above. The large linewidth of $\Delta B=48$~mT indicates the presence of comparably fast relaxation processes in \LiMnSO\ which is straightforwardly explained by residual short-range magnetic correlations at 300~K. We conclude that, at 300~K, \LiMnSO\ is not yet in the purely paramagnetic regime but exhibits sizable magnetic correlations.}

Considering that $z$ is independent on the transition metals in \LiMChO , the drastic increase of $3T_{\rm N}/(S(S+1))$ suggests a strong increase of $|J_{\rm eff}|$ under the exchange Mn $\rightarrow$ Fe $\rightarrow$ Co at the $M$ site (see Table~\ref{tab:structure}). Alternatively, one may speculate that, due to the vicinity to the percolation threshold, the variation in the magnetic anisotropy from the almost isotropic spin systems \LiMnChO\ to the strongly anisotropic ones in \LiCoChO\ affects the value of \tn . {\rev The transition metal ions under study are paradigmatic examples of vanishing (Mn$^{2+}: 3d^5; J=S=5/2)$, intermediate (Fe$^{2+}: 3d^6$) and strong (Co$^{2+}: 3d^7$) magnetic anisotropy in their high-spin states realized at hand.~\cite{AbragamBleaney, Moseley2022} For \LiMnChO , we have confirmed the expected vanishing anisotropy by means of ESR which shows $g\approx 2$. As expected for ions with pronounced magnetic anisotropy, we did not find X-band ESR signals in the Fe-based materials which are usually ESR-silent at X-band frequencies. Future studies for the quantitative determination of the expected increase of magnetic anisotropy when replacing Fe by Co, e.g., by inelastic neutron scattering, X-ray magnetic circular dichroism, or magnetostriction measurements in combination with numerical studies will be needed to further disentangle the magnetic and structural parameters governing magnetism in Li-rich antiperovskites.}

{A key effect of magnetic anisotropy on the actual ordering temperatures} is suggested by the {\rev reported} different critical behavior of relevant properties such as the magnetic correlation length {\rev in materials with different magnetic anisotropy}.~\cite{Birgeneau1984,Ballesteros1998} {\rev Furthermore, different dilution dependencies of the actual ordering temperatures have been observed} when approaching the transition temperature. The fact that \tn\ is clearly larger in \LiCoSeO\ than in \LiCoSO\ then suggests that exchanging S by Se affects either $|J_{\rm eff}|$ or the magnetic anisotropy in \LiCoChO . In contrast, this effect is not observed in the Fe and Mn-based materials (see Fig.~\ref{fig:tn}). 

In addition to long-range magnetic order, %, which is unambiguously proven via Mössbauer spectroscopy by a vanishing hyperfine field at \tn\ $\simeq 50$~K , 
reported NMR data reveal short-range magnetic order persisting up to about 100~K in \LiFeChO .~\cite{Seewald2025} Notably, this coincides with the onset temperature of ZFC/FC bifurcation in \LiFeChO\ which is referred to as \tirr . We infer the presence of uncompensated magnetic moments in the short-range ordered phase which is reflected by a weak canted moment appearing below \tn . Qualitatively, FC/ZFC bifurcation is a typical feature in strongly diluted AFM systems in the vicinity but above the percolation threshold.~\cite{marinari2000,shand1998,martinho2001,kemei2014}. This also holds for the presence of a small canted moment and the appearance of weak ferromagnetism since quenched disorder leads to spatially inhomogeneous magnetic exchange interactions. In such diluted systems, rare regions with strong magnetic coupling, embedded in a paramagnetic background, can maintain finite localized magnetic moments even above the ordering temperature~\cite{rieger1996}. Finally, we note that unusual magnetic phases, including the Griffiths phase found by means of power law analysis of the magnetic susceptibility in several diluted magnets~\cite{li2024,nair2011}, are not indicated in the materials studied at hand (see Fig.~\ref{powerlaw} in the Appendix).

\subsubsection*{Itinerant electron scenarios}

The observation of an only weakly temperature-dependent magnetic susceptibility motivates to discuss the scenario of weakly itinerant electrons. Indeed, metallic systems with rather poor conductivity can show intriguing electronic instabilities. The fragile nature and strong air-sensitivity of Li-rich antiperovskites challenges electronic conductivity measurements and, to the best of our knowledge, its temperature dependence has not been studied yet. At room temperature, direct current polarization studies of the electronic conductivity of \lifeso\ on pressed pellets using ion-blocking electrodes showed $\sim 5-8\times 10^{-4}$~S/cm which is in the regime of semiconducting behavior~\cite{Hikima2024}. {\rk Possible lithium deficiency and associated oxidation of Fe$^{2+}$ can however shift the Fermi level deeper into the Fe 3$d$ manifold so that partially filled $d$ states yield greatly improving electronic conductivity.}~\cite{Gao2023}

Typical Pauli susceptibilities of metals are in the order $10^{-4}-10^{-5}$~\ergGmol\ which is much smaller than the susceptibilities observed at hand. This agrees to the following estimate where the DOS of $D(E_{\rm F})\simeq12~\mathrm{Ry}^{-1}$ per unit cell numerically obtained in~\cite{zarhri2024ab} is used. For non-correlated metals, this yields $\chi_\mathrm{Pauli} = N_{\rm A}\mu_{\rm B}^2\times D(E_{\rm F})/Z = 1.4 \times 10^{-5}$~\ergGmol. Here, $N{\rm _A}$ is Avogadro's number and $Z = 2$ is the number of formula units in the crystallographic unit cell. This value of $\chi_\mathrm{Pauli}$ is in the regime expected for typical non-correlated metals but is two orders of magnitude smaller than the experimentally derived high-field susceptibilities shown in Tab.~\ref{tab:tn}. When attributing this discrepancy to a Stoner-like enhancement of the magnetic susceptibility as described by 
%\begin{equation}
$\chi_{\text{enh}} = \chi_{\mathrm{P}}/(1 - I D(E_{\mathrm{F}}))$
%\end{equation}
with $I$ being the exchange integral describing the magnetic exchange interaction, our experimental data would indicate $ID(E_{\rm F})\simeq 0.99$. Such a large Stoner factor would place the antiperovskites under study at the verge of long-range FM order and is rather unlikely. 

%The lack of reliable transport studies on \LiMChO\ hinders the interpretation of the magnetic data as 
While our data exclude non-correlated models, a variety of correlated metals are reported in the literature featuring enhanced Pauli-like susceptibility. In CoS$_{1.4}$Se${_{0.6}}$, Goto \etal\ report $\chi \simeq 2\times 10^{-3}$~\ergGmol\ which is of similar order as found at hand~\cite{goto1997} and has been attributed to zero-point spin fluctuations. 
Strong Pauli-enhancement with a Stoner enhancement factor of $\sim 20$ is reported for the cubic Cr$_{23}$B$_6$-type compounds Lu$_{2-x}$Ni$_{21}$B$_6$ and Zr$_2$Ni$_{21}$B$_6$ which show $\chi=1.7\times 10^{-3}$~\ergGmol .  
Rather large magnetic susceptibility has been also found in pnictide-based superconductors at temperatures above the spin-density-wave transition~\cite{klingeler2010}. Different to our findings at hand, it features an increase of $\chi$ upon heating. Among a variety of possible scenarios, it has been argued that the magnetic susceptibility in Fe-based pnictides is linked with a sharp peak in the spectral function located slightly below the Fermi level.~\cite{Vollhardt2012}. Finally, we also note that the disordered nature of \LiMChO\ may introduce a series of impurity levels which also might strongly modify the electronic response as discussed, e.g., for W$_{18}$O$_{49}$~\cite{Wang2016}. {\rk We therefore conclude that, although we consider an itinerant electron scenario rather unlikely, possible electronic itinerancy induced by lithium deficiency cannot be completely ruled out based on our magnetic measurements, but would necessarily imply the presence of strong correlation effects.}

\section{Conclusions}

The lithium-rich antiperovskites \LiMChO\ ($M$ = Fe, Co, Mn and $Ch$ = Se, S) enable us to investigate a series of diluted magnetic systems with identical dilution levels just above the percolation threshold but variation of spin size, magnetic anisotropy, and magnetic coupling. In the antiperovskite \abx\ structure, transition metal and lithium ions are randomly distributed on the X-sites of the \abx\ structure, thereby forming an 2/3-diluted (i.e., 1/3-filled) magnetic lattice. While all materials under study show long-range magnetic order, the ordering temperatures strongly increase from $\simeq 30$~K in \LiMnChO\ to $\simeq 50$~K in \LiFeChO\ and $70-90$~K in \LiCoChO . We conclude significant magnetic coupling ($|zJ_{\rm eff}|>100$~K) and attribute the transition-metal dependence of \tn\ to an increasing magnetic anisotropy of the magnetic subsystems. At high temperatures, the data imply the absence of a distinct Curie-Weiss behavior below 350~K but show a large Pauli-like magnetic susceptibility in the order of $\chi \simeq 10^{-3}$~\ergGmol\ in all materials under study. Itinerant electron scenarios are rather unlikely. Instead, we ascribe the temperature dependence of $\chi$ to the response of diluted, strongly interacting but only short-range ordered magnetic systems. 

%Lu et al. calculated the density of states of Li2FeSO at the GGA + U level. The material is computed to be semi-conducting with a band gap of ~2.7 eV. The magnetic moment is calculated as 4 \mbfe .~\cite{lu2018} In \LiCoSO\ and \LiMnSO , the band gap is calulated to about 3 eV and 2 eV, respectively.

%Metallic behaviour in \lifeso\ suggested here (but not discussed): https://doi.org/10.1051/e3sconf/202458201002

%\LiFeSeO\ and \LiMnSO\ are predicted to be semimetals with a large bandgap from first-principle calculations here: https://doi.org/10.1016/j.cclet.2024.110683 . A transition to a metallic state is predicted by doping with elements such as Na, K, F, and Cl. This is due to the addition/removal of valence electrons in the system
%after doping with these elements, which causes the Fermi level to shift upward/downward.

%Gorbunov: \cite{Gorbunov.2024} Our analysis shows that for \LiMSeO\ with M = Mn, Co, Fe, the lowest energy
%configuration had zero net moment. For all the compounds, we obtained an insulating solution with a band gap ranging from 1.5 to 2 eV. 

%%%%%%%%%%%%%%%%%%%

\begin{acknowledgements}
We thank Florin Krasniqi (U Heidelberg) for technical support. Work has been done in the framework of the joint project KL 1824/20-1 \& GR 5987/2-1 funded by Deutsche Forschungsgemeinschaft (DFG). J.Z. and L.B. acknowledge the DFG Research Training Group “Mixed Ionic Electronic Transport” (GRK 2948). J.~Zheng acknowledges fellowship by the Chinese Scholarship Council (File No. 202304910069). Work has also been supported within the framework of the Excellence Strategy of the Federal and State Governments of Germany via the Heidelberg University's flagship EMS initiative and the Cluster of Excellence STRUCTURES, and via the TU Dresden/U Würzburg Cluster of Excellence ct.qmat. M.A.A.~M. thanks the IFW excellence program for financial support. 
\end{acknowledgements}

\bibliography{ref}

%\hfill
%\newpage

%\onecolumngrid
%\newpage 

%\begin{center}
%{\LARGE Supplemental Material:\\ Magnetism in a series of antiperovskite (Li$_2$\textit{M})\textit{Ch}O (\textit{M} = Fe, Mn, Co; \textit{Ch} = S, Se) diluted magnets with fixed 1/3 filling ratio}\

\section*{Appendix}
\label{sec:appendix}
%\vspace{0.5cm}

%{\large J. Zheng$^1$, F. L. Carstens$^1$, L. Singer$^1$, M.A.A. Mohamed$^{2,3}$, J. Arneth$^1$, S. Hampel$^2$, N. Gräßler$^2$, R. Klingeler$^1$}

%\vspace{0.3cm}

%$^1$Kirchhoff Institute for Physics, Heidelberg University, INF 227, D-69120 Heidelberg, Germany

%$^2$Leibniz Institute for Solid State and Materials Research Dresden, 01069 Dresden, Germany

%$^3$Department of Physics, Faculty of Science, Sohag University, 82524 Sohag, Egypt
%\end{center}

\renewcommand{\thefigure}{A\arabic{figure}}
\setcounter{figure}{0} 
\setcounter{section}{0}

\renewcommand{\thetable}{A\arabic{table}}
\setcounter{table}{0} 
\setcounter{section}{0}

\renewcommand{\topfraction}{1}

%The Supplemental Material contains further information on:

%- The mechanochemical synthesis and the characterization of the materials by SEM, and ICP-OES.

%- The nature and amount of impurity phases as deduced from the magnetization studies.

%- The potential impact of impurity phases on the experimentally determined high-field susceptibility.

%- Numerical data on the magnetic response of finite spin ensembles.
    
%- The assessment of potential power-law behavior of the magnetic susceptibility.

\section{Mechanochemical Synthesis}

All materials under study have been made by mechanochemical syntheses which provides a simple, easily scalable, and laboratory-safe process. 
For $M$ = Fe the synthesis method as well as the effects of post-synthesis heat-treatment to eliminate and to alter various impurity phases %which inevitably appear in this class of materials 
have been reported in our previous works~\cite{Mohamed2023mechanochemical,Singer2023elucidating,Singer2024,Mohamed2023}; here, we extend this approach to the materials \LiMChO\ with $M$ = Co and Mn.

The samples of \LiMChO\  ($M$ = Co, Mn; $Ch$ = S, Se) were synthesized by high-energy ball milling using stainless-steel jars. All sample handling and loading were performed inside an argon-filled glovebox (MBraun) with controlled atmosphere (O$_2$ and H$_2$O $< 1$~ppm). For the sulfur-containing compounds, (Li$_2$M)SO ($M$ = Co, Mn), stoichiometric amounts of Li$_2$S and the corresponding transition-metal oxide (CoO or MnO) were used as starting materials. For the selenium-containing compounds, (Li$_2$M)SeO (M = Co, Mn), stoichiometric mixtures of Li$_2$O, elemental M (Co or Mn), and elemental Se were employed. The total mass of the starting mixtures ranged between 4 and 6~g. The ball-to-powder weight ratio (BPR) was maintained at 5:1 for all compositions, except for \LiCoSeO , for which a BPR of 3:1 was employed. High-energy ball milling was carried out at room temperature using a SPEX SamplePrep 8000D mixer/mill operated at a rotation speed of 875~rpm. The milling process was interrupted at regular intervals, and the reaction progress was monitored by %powder x-ray diffraction 
pXRD. Milling was continued until no further changes in the diffraction patterns were observed, indicating completion of the reaction. At the end of each milling step, the jar was opened inside the glovebox, and the powder was extracted for subsequent characterization. 

%\begin{figure}[h]
%\begin{center}
% \includegraphics[width=0.7\columnwidth,clip]{XRD for all six.png}
% \caption{X-ray diffraction patterns of the investigated samples. The horizontal axis represents the scattering vector $Q$ (instead of 2$\theta$) because the measurements were performed by using different X-ray wavelengths. The dark cyan lines in each plots mark the Bragg positions of the expected main phases. Impurity phases are only detected in \LiMnSeO\ as shown by the markers. The data of \LiFeSO\ and \LiFeSeO\ are reprinted from Ref.~\cite{Mohamed2023,Singer2023elucidating}.}
%\label{fig:xrdappendix}
%\end{center}
%\end{figure}

\begin{figure}[h]
\begin{center}
 \includegraphics[width=1\columnwidth,clip]{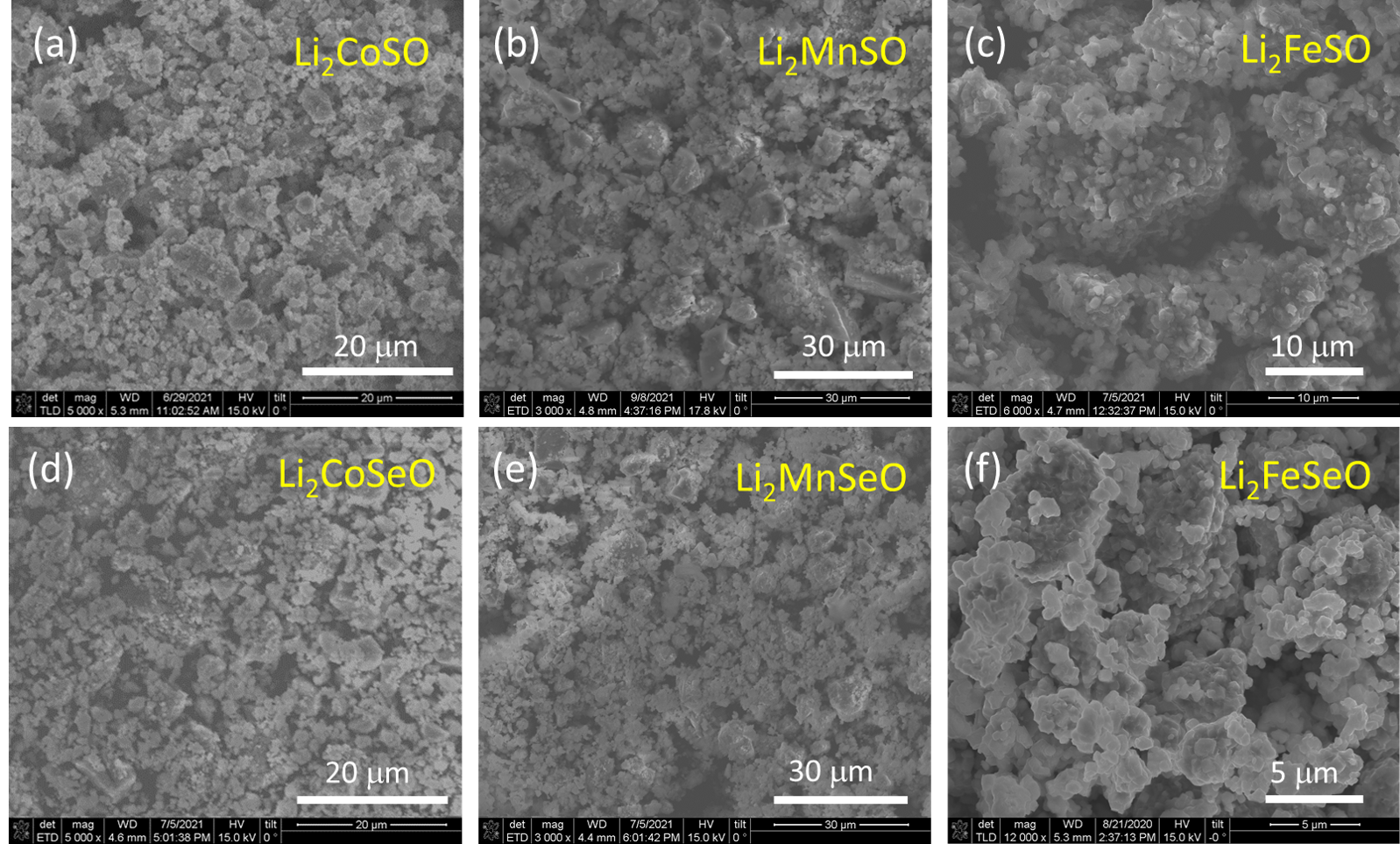}
 \caption{SEM images for the studied \LiMChO\  ($M$ = Fe, Co, Mn; $Ch$ = S, Se) materials.}
\label{fig:sem}
\end{center}
\end{figure}

\begin{table*}[htb]
    \centering
    \begin{tabular}{l|l|l}
    \hline\hline
        \multicolumn{3}{c}{ICP-OES data}  \\ \hline
         &Mass-percent&Molar ratios\\ \hline
        \LiMnSO & 
        Li$_{11.74(23)}$Mn$_{46.17(87)}$Fe$_{1.39(3)}$S$_{26.0(7)}$&
        Li$_{1.97(4)}$Mn$_{0.98(2)}$Fe$_{0.03(1)}$S$_{0.95(3)}$O$_{1.08}$ \\
        \LiMnSeO & Li$_{7.86(6)}$Mn$_{31.22(31)}$Fe$_{4.54(7)}$Se$_{45.99(48)}$ & Li$_{1.88(2)}$Mn$_{0.94(1)}$Fe$_{0.13(1)}$Se$_{0.97(1)}$O$_{1.08}$ \\
        \LiFeSO & Li$_{11.52(9)}$Fe$_{46.90(2.20)}$S$_{25.9(5)}$&Li$_{1.93(2)}$Fe$_{0.98(5)}$S$_{0.94(2)}$O$_{1.15}$ \\
        \LiFeSeO & Li$_{8,36(8)}$Fe$_{33.81(28)}$Se$_{47.49(49)}$&Li$_{1.99(2)}$Fe$_{1.01(1)}$Se$_{1.00(1)}$O$_{0.99}$ \\
        \LiCoSO & Li$_{11.29(26)}$Co$_{48.86(93)}$S$_{25.48(42)}$ &Li$_{1.96(4)}$Co$_{0.96(1)}$S$_{0.96(2)}$O$_{1.08}$ \\
        \LiCoSeO & Li$_{7.70(8)}$Co$_{33.52(2)}$Fe$_{2.88(2)}$Se$_{45.68(67)}$ &Li$_{1.98(2)}$Co$_{0.96(1)}$Fe$_{0.09(1)}$Se$_{0.98(2)}$O$_{1.08}$ \\
    \hline\hline
    \end{tabular}
    \caption{Composition of the studied materials as determined by element analysis through ICP-OES measurements. The molar ratio was calculated from the obtained mass-percentages of the ICP-OES analyses. Molar ratios are scaled to sum up to about 5.}
    \label{tab:ICPOES}
\end{table*}

%The XRD patterns of the samples under study are presented in Fig.~\ref{fig:xrdappendix}. The data confirm the main \LiMChO\ phases. Depending on the composition, different tiny amounts of impurity phases are detected. The XRD results shown in Fig.~\ref{fig:xrdappendix} are from pristine samples without heat-treatment. 

%Figure \ref{fig:xrdappendix} presents the pXRD patterns of the as-ball-milled samples. All target phases were successfully synthesized, as evidenced by diffraction patterns exhibiting broad but well-defined Bragg peaks that are correctly positioned and match the reported ICSD/CSD reference patterns for the corresponding compounds. The significant peak broadening indicates small crystallite sizes, consistent with high-energy mechanochemical synthesis. The detailed discussion of the impurity phases in Li$_2$Fe$Ch$O can be found in our previous works Refs.~\cite{Mohamed2023mechanochemical,Singer2023elucidating,Seewald2025}. Minor crystalline secondary phases were only detected for \LiMnSeO .

In addition to the XRD data presented in the main manuscript, the successful synthesis was further confirmed by inductively coupled plasma optical emission spectroscopy (ICP-OES), as shown in Table~\ref{tab:ICPOES}. The measured compositions are in good agreement with the nominal stoichiometries. Minor Fe contamination was detected in most samples, particularly in the Se-based compounds, and is attributed to abrasion of the milling balls and vial during high-energy dry milling. Scanning electron microscopy (SEM) images of the ball-milled samples (Fig.~\ref{fig:sem}) reveal predominantly spherical nanoparticles with micrometer dimensions and broad size distribution.

%Original text by RK: ICPOES measurements were conducted to analyze the composition the Mn- and Co-based samples. Notably, Fe was detected in \LiCoSeO , \LiMnSO  and \LiMnSeO. We attribute this to Fe from the ball-milling process due to the usage of stainless-steel beads. ICPOES implies $\sim 1.4$ w\% and $\sim 4.7$ w\% of Fe in \LiMnSO  and \LiMnSeO, respectively. The Fe signal is also observed in \LiCoSeO, which yields $\sim 2.9$w\% Fe in \LiCoSeO. ICPOES has not detected any external element in \LiCoSO.  

\section{Nature and amount of magnetic impurity phases}

\begin{figure} [htb] 
    \includegraphics[width = 0.9\columnwidth]{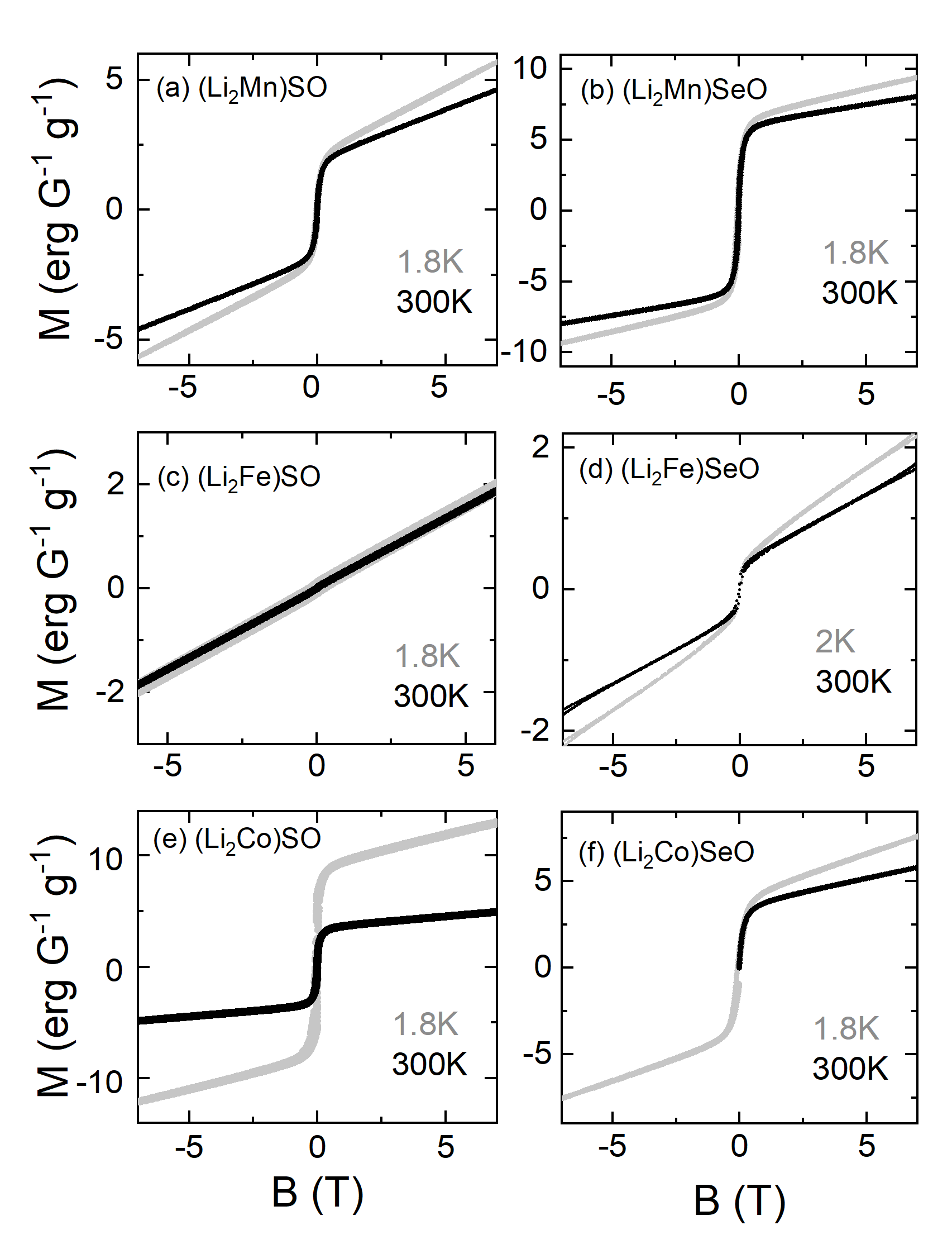}
    \caption{Isothermal magnetization of \LiMChO\ with $M$ = Mn, Fe, Co and $Ch$ = S, Se as indicated.
    }
    \label{MB}
\end{figure}

Fig~\ref{MB} shows isothermal magnetization curves of the materials under study. All specimens exhibit small FM impurity signals at low magnetic field, except for \LiFeSO . The saturation magnetization $M_{\rm sat}^{\rm FM}$ of the FM impurities is extracted from the isothermal magnetization curves and shown in Table~\ref{tab:impurities}. The impurities in the heat-treated Fe-based samples are most likely Fe$_{1-x}$Se and Fe$_x$S, respectively.%, while metallic Fe present in the pristine materials vanished upon the heat-treatment
~\cite{Singer2024,Singer2023elucidating}.

\begin{table*}[htb]
    \centering
    \begin{tabular}{l|c|c|c|c}
    \hline\hline
         & $M_{\rm sat}^{\rm FM}$ at 300~K & pXRD & ICP-OES & Conclusion \\
         \hline
       \LiMnSO & 2~erg/(Gg) & - & 1.39(5)~w\% Fe & Fe$^0$ \\
       \LiMnSeO & 6~erg/(Gg)& MnO$_2$, SeO$_3$ & 4.54(7)~w\% Fe & Fe$^0$ and Fe$_3$O$_4$ \\
       \LiFeSO & 0.03~erg/(Gg) & Fe$_x$S & - & $\sim$ 0.2~w\%\ Fe$_x$S ~\cite{Singer2024} \\
       \LiFeSeO & $<1$ erg/(Gg) & - & - & $<0.1$\% Fe$_{1-x}$Se~\cite{Singer2023elucidating} \\
       \LiCoSO & 3.5~erg/(Gg) & -\footnote{Co$_9$S$_8$ appears after heat-treatment.}& - & Co$^0$ \\
       \LiCoSeO & 3.5~erg/(Gg)& -\footnote{Co$_9$Se$_8$ appears after heat-treatment.}   & 2.88(2)~w\%Fe & Fe$^0$, Co$^0$, Fe$_3$O$_4$ \\ 
    \hline\hline
    \end{tabular}
    \caption{Summary of characterizing studies on \LiMChO\ studied at hand. $M_{\rm sat}^{\rm FM}$ is the size of the temperature-independent FM moment. The rows labeled pXRD and ICPEOS summarize the results of the respective measurement. No value indicates the absence of pXRD features which implies the absence of more than 1-2~\% of crystalline impurities. The last row summarizes the conclusions on the magnetic impurity phases.}
    \label{tab:impurities}
\end{table*}

 %Notably, for the Fe-based samples measured in this study additional heat treatment reduces the FM impurities (i.e., metallic Fe in the pristine samples) as reported previously~\cite{Singer2024,Singer2023elucidating}. The Fe-based samples feature the smallest FM impurity contents in the series under studies, with \LiFeSO\ showing nearly no traces of such phases. 

For the Mn-containing compositions, the FM signals can neither be metallic Mn nor its oxides/chalkogenides because they are AFM at room temperature. In contrast, the ICP-OES data suggest a tiny Fe signal which likely originates from the ball-milling process. The FM impurities can hence be in principle attributed to metallic Fe$^0$ or Fe$_3$O$_4$. By using the saturation magnetization of Fe$^0$ ($\sim 222$ erg/(Gg)) and Fe$_3$O$_4$ ($\sim 80$ erg/(Gg))~\cite{nguyen2021} hence the amount of both potential impurity phase can be estimated by the experimentally determined $M_{\rm sat}^{\rm FM}$. For \LiMnSO , this analysis yields $~\sim 0.9$~w\% for Fe$^0$ or $~\sim 2.5$~w\% for Fe$_3$O$_4$, and $\sim 2.7$~w\% Fe$^0$ or $\sim7.5$~w\% Fe$_3$O$_4$ for \LiMnSeO. We conclude the presence of Fe$^0$ in \LiMnSO\ while both Fe$^0$ and Fe$_3$O$_4$ can be present in \LiMnSeO . The fact that the Verwey-transition is not seen in the data implies that, if Fe$_3$O$_4$ is present, it would have to be nanosized~\cite{Gautam2026}.

%We calculate that \LiMnSO  contains $\sim 0.4$ w\% metallic Fe and $\sim 1.4$ w\% Fe$_3$O$_4$, and \LiMnSeO  contains$\sim 0.7$ w\% metallic Fe and $\sim 5.5$ w\% Fe$_3$O$_4$ .

In the Co-based materials, metallic Co must be considered as potential FM impurity. The origin of ferromagnetism is hence less clear than in the Fe- and Mn-counterparts if only the magnetization data are considered: It may be metallic Co$^0$ (with a saturation moment of $\sim 161$~erg/(Gg)), metallic iron, or iron oxide -- or a combination of them. For metallic Co, $M_{\rm sat}^{\rm FM}$ suggests an impurity content of approximately 2.2\% Co$^0$ in both \LiCoSO\  and \LiCoSeO . We note that we exclude the presence of CoS$_2$ in \LiCoSO\  since this material is FM only below 112~K (typical $\sim 38$ erg/(Gg)). %The magnetization of \LiCoSO\ at 1~T also shows a more pronounced temperature dependence. 
Both the absence of a clear feature at 112~K (see Fig.~\ref{SQUIDall}) as well as the high amount of $\sim 20$\% CoS$_2$ needed to account for the observed saturation moment excludes this scenario because such high amounts of impurity phases would be detected in XRD.

Considering the limitations on the determination of O, and to some extend on S and Se, by means of ICP-OES, in combination with the magnetization data we conclude the following picture: Firstly, the ICP-OES Fe signal cannot be totally due to metallic Fe$^0$ in \LiMnSeO . The ICP-OES result (1.4~w\% Fe$^0$ in \LiMnSO , 4.5~w\% in \LiMnSeO\ and 2.9~w\% in \LiCoSeO ) agrees reasonably well with the magnetization data only for \LiMnSO\ but would lead too large magnetic moments for the latter two materials. If one instead assumes that the Fe signals entirely originate from Fe$_3$O$_4$, ICP-OES would suggest 2~w\% of Fe$_3$O$_4$ in \LiMnSO , 6.1~w\% in \LiMnSeO , and 4.1~ w\% in \LiCoSeO\ which yields lower remanent magnetization than experimentally observed. We conclude for the latter two materials that the Fe signal originates from a mixture of Fe$^0$ and Fe$_3$O$_4$. Our data do not exclude the presence of non-magnetic iron oxide. In \LiCoSO , ICP-OES does not report any Fe signal and metallic Co$^0$ is considered the only FM impurity phase at room temperature. In \LiCoChO~and \LiMnSeO , ICP-OES signals slight transition-metal-deficiency which indicates potential metal vacancies in the main phase. %, i.e., V$\rm_{Mn}$ and V$\rm_{Co}$. 
Typically, such defects can contribute to the magnetic response by a Curie-like signal which is however not observed in our magnetic susceptibility data. %We note, that the data do not exclude the presence of elemental transition metal, i.e., Co$^0$ in \LiCoChO . 
The nature and quantities of the impurities as deduced from the detailed discussion above are summarized in table~\ref{tab:impurities}.

%{\jy Attributing the Fe signal to Fe$^0$:for \LiMnSO, Fe$^0$: 0.9\% from M(B) vs. 1.4\% from ICP. The derivation occurs in \LiMnSeO, Fe$^0$: 2.7\% from M(B) and 4.7\% from ICP.For \LiCoSeO\ ICP detect Fe signal and 2.9\% Fe$^0$ is estimated. M(B) suggests 1.6\% Fe$^0$.For \LiCoSO\ ICP did not detect Fe impurity, and did not support the existence of Co$^0$ (average valence state of Co$^{+2.23}$). However, if the impurity is indeed the Co$^0$. M(B) gives 2.2\% Co$^0$ (Otherwise probably no magnetic impurity as the candidate, no idea why ICP did not support this scenario). Notably, the ICP of \LiCoSeO also did not support the emergence of Co$^0$. If we consider the Fe signal from the ICP is totally Fe$_3$O$_4$. For \LiMnSO, ICP suggests 1.98\% Fe$_3$O$_4$, M(B) suggests 2.5\%. For \LiMnSeO, ICP suggests 6.05\% and M(B) suggests 7.5\%. The morphology of Fe$_3$O$_4$ affects its saturation moment. Since we do not know the morphology, the arbitrary value of 80\ergGg\ may lead to the discrepancy.For \LiCoSeO, ICP suggests 4.09\% Fe$_3$O$_4$, and M(B) suggests 4.38\% Fe$_3$O$_4$.}

%\subsection{Potential impact of impurity phases on the high-field susceptibility}

By tracing the nature and quantities of the impurity phases, we can further assess their impact on the obtained high-field differential susceptibilities presented in the main text. Although certain impurities such as metallic Fe$^0$, Co$^0$, and Fe$_{1-x}$Se exhibit relatively high intrinsic magnetic susceptibilities (on the order of 
$\sim 10^{-4}$ \ergGmol)~\cite{herring1966,rebouillat2003,mankovsky2006,hirone1954}, their contribution to the overall susceptibility of the sample is minimal. Given their dilute concentrations within the host phases, the estimated impurity phase contribution $\chi_{\rm imp}\sim 10^{-6}$\ergGmol\  is rather small. Furthermore, Fe$_3$O$_4$ and Fe$_x$S have a negligible impact because they are magnetically saturated under the applied field of 6~T\cite{kind2013,volk2016,ozkaya2009,kihal2012}. Although a recent report suggested a high-field susceptibility for 3c Fe$_x$S as large as $\sim 10^{-3}$ \ergGmol~\cite{jyotsna2023}, its exceedingly low concentration in the host phases ensures that its net contribution remains $\chi_{\rm imp}\sim 10^{-6}$~\ergGmol .

In light of the data presented in Table~\ref{tab:tn}, we argue that the influence of these impurities on the high-field Pauli-like susceptibility is negligible. Consequently, the 
$\chi_0$ value extracted from the high-field region (6 T) of the M(B) curves can be reliably attributed to the intrinsic magnetic properties of the antiperovskite host phase.

%{\rk The next paragraph is disordered and nearly impossible to understand. REVISE! the  Fe$^0$, Co$^0$ or Fe$_x$S, despite possessing the rather high susceptibilities ($\sim 10^{-4}$\ergGmol) among the list of impurities\cite{herring1966,rebouillat2003,mankovsky2006,hirone1954}, contribute only minimally to the overall magnetic susceptibility. According to their concentration in the host phases, they can only contribute $\chi_{\rm imp.}\sim 10^{-6}$\ergGmol\ to total susceptibility. Fe$_3$O$_4$ and Fe$_x$S have negligible impact since both are saturated at 6T\cite{kind2013,volk2016,ozkaya2009,kihal2012}. In particular, although few reports indicate that the high field susceptibility of 3c Fe$_x$S is as high as $\sim 10^{-3}$\ergGmol\cite{jyotsna2023}, its extremely low content in the host phases means its overall contribution remains on the order of $\sim 10^{-6}$. Comparing the data presented in Table~\ref{tab:tn}, we argue that the impact of the impurities on the high field Pauli-like susceptibility is negligible. The $\chi_0$ extracted from the M(B) at 6T should be attributed solely to the intrinsic magnetic property of the antiperovskite host phases.}

%\newpage 

\section{Magnetic response of finite spin ensembles}

\begin{figure}[tb]
\begin{center}
\includegraphics[width=0.99\columnwidth,clip]{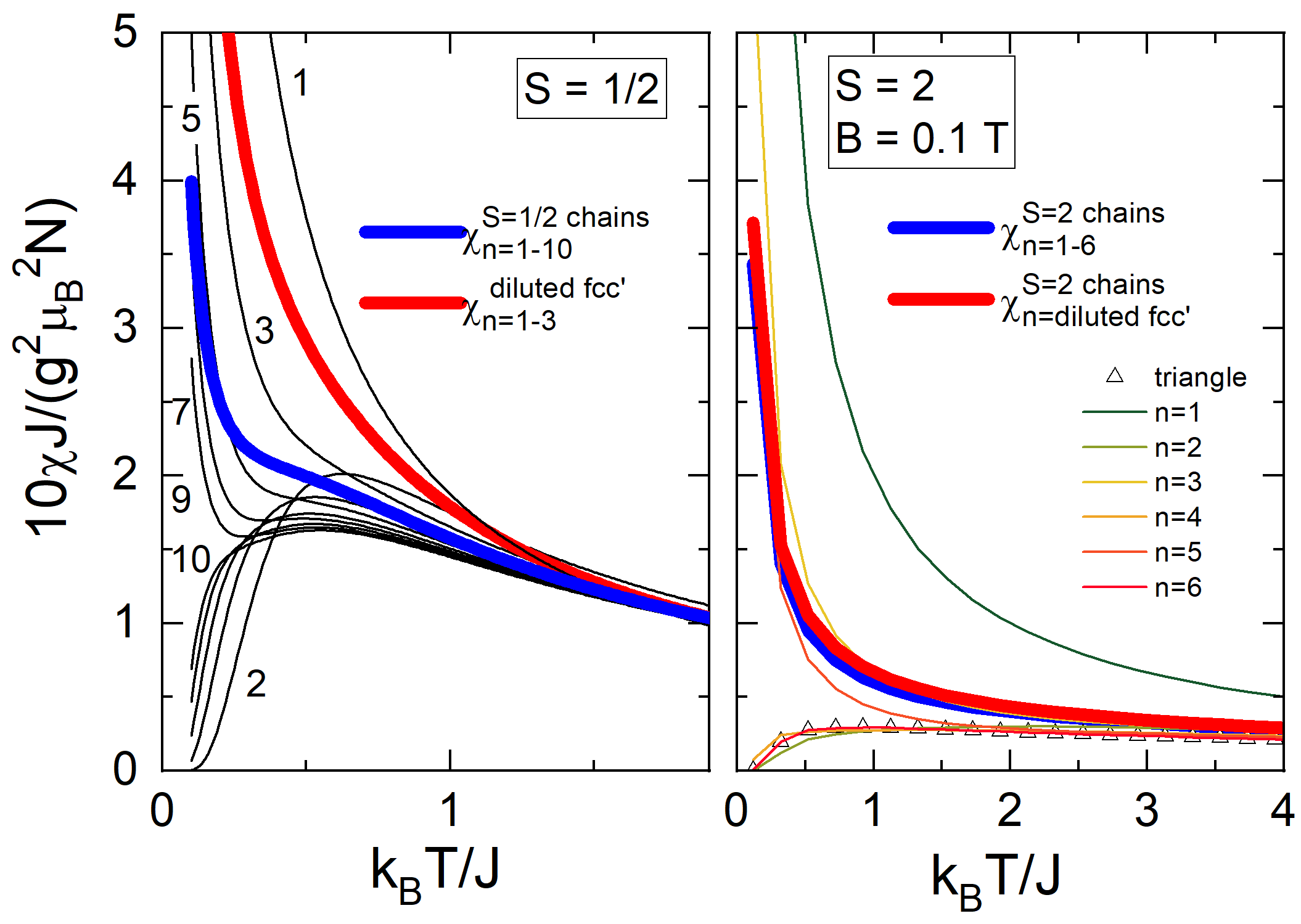}
 \caption{Numerical results on the magnetic response of finite spin chains $S=1/2$ ($n\leq 10$) and $S=2$ ($n\leq 6$) and $S=2$ triangles. Blue lines show the summarized response of evenly distributed finite chains; red lines estimate the response by using approximations for the probability of finite clusters as given in the main manuscript text.}
\label{fig:cluster}
\end{center}
\end{figure}

In principle, the superimposed magnetic response of isolated finite spin ensembles may result in a weakly temperature dependent magnetic susceptibility.~\cite{Garcia2015,reis2006,deisenhofer2006} Two illustrative examples of these superimposed responses are shown in Fig.~\ref{fig:cluster}. In the figure, the responses of finite spin chains with $S=1/2$ and $S=2$ are shown, together with the summarized response of evenly distributed finite chains. Relevant here is the superimposed response of chain fragments appearing with the probabilities calculated in the main text for the diluted incomplete fcc lattice (diluted fcc') as shown by the red lines. The small probability of finding isolated small clusters with $n\geq 2$ rules out the scenario of isolated small clusters. However, including next nearest interaction and frustration to the disordered percolating network of isolated spins may  qualitatively explain the observed temperature dependence of the magnetic susceptibility as described in the main text.

\section{Potential power-law behavior of the magnetic susceptibility}

\begin{figure}[tb]
\begin{center}
  \includegraphics[width=0.95\linewidth]{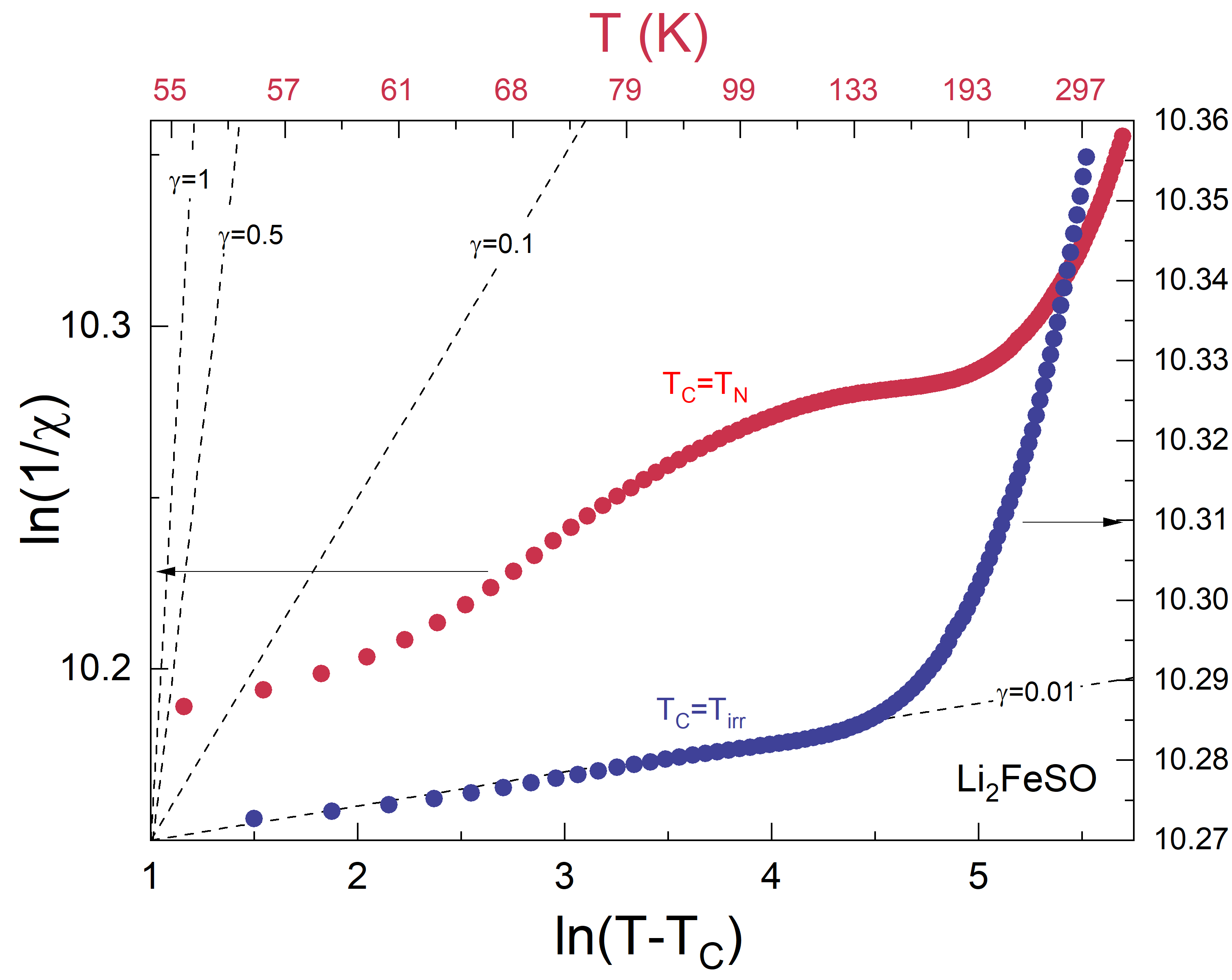}
    \caption{Double logarithmic plot of $\chi^{-1}$ vs. reduced temperature ($T-T_{\rm C}$) of Li$_2$FeSO. For the critical temperature, both the actual magnetic ordering temperature \tn\ and the irreversibility temperature \tirr\ have been chosen (i.e., the temperature below which there is a thermal FC/ZFC hysteresis; see Fig.~\ref{SQUIDall} in the main manuscript text). Dashed lines shows the hypothetical power law behavior associated with different critical exponents $\gamma$. The upper temperature scale refers only to the red data points.}
\label{powerlaw}
\end{center}
\end{figure}

Figure~\ref{powerlaw} shows a double logarithmic plot of $\chi^{-1}$ vs. the reduced temperature ($T-T_{\rm C}$) of Li$_2$FeSO in order to investigate potential critical behavior $\chi\propto (T-T_{\rm C})^{-\gamma}$. For the critical temperature, both the actual magnetic ordering temperature \tn\ and the irreversibility temperature \tirr\ have been chosen. In a model of rare regions in a diluted magnet in the paramagnetic phase, \tc\ is the critical temperature of random FM clusters. In ferromagnets, this temperature can be either the Curie temperature or the irreversibility temperature. For antiferromagnets with Griffiths-like behavior, \tc\ is usually selected among \tn , \tirr , the Weiss temperature or a free fitting parameter~\cite{ouyang2011,guo2008}. Here, no Curie-Weiss behavior is observed in the accessible temperature regime so that we choose $T_{\rm C}=T_{\rm N}$ and $T_{\rm C}=T_{\rm irr}$, respectively. The data in Fig.~\ref{powerlaw} however show the absence of critical scaling. We also note that it further corroborates the non-mean-field behavior of the magnetic susceptibility up to 350~K.

\end{document}